\documentclass[aps,prl,amsmath,amssymb,reprint,floatfix]{revtex4-1}
\usepackage[utf8]{inputenc}
\usepackage{mfirstuc}
\usepackage{graphicx}
\usepackage{subfigure}
\usepackage{float}
\usepackage{gensymb}
\usepackage{amsmath}
\usepackage{color}
\usepackage{multirow}
\usepackage{natbib}
\usepackage{comment}
\usepackage[colorlinks=true, linkcolor=blue, citecolor=blue, urlcolor=blue]{hyperref}
\usepackage{breakurl}
\usepackage{url}
\newcommand{\SK}[1]{\textcolor{black}{{#1}}}

\begin{document}
%\begin{titlepage}

\title{Understanding the Role of Particle Deformability on the Crystal and Glass formation using Two-dimensional Ring Polymer Model}
\author{Padmanabha Bose}
\author{Smarajit Karmakar}
\email{smarajit@tifrh.res.in}
\affiliation{Tata Institute of Fundamental Research, 36/P, Gopanpally Village, Serilingampally Mandal, Ranga Reddy District, Hyderabad 500046, Telangana, India}

%\maketitle

\begin{abstract}
\SK{Soft matter systems are common in nature and make up nearly all the essential components necessary for life, from cells to the organelles within those cells. The ability of these soft materials to deform is crucial for the proper functioning of various biological processes, such as blood flow in our veins and arteries. It is vital to understand how deformability influences the normal functioning of these processes. We have investigated an assembly of two-dimensional (2D) polymeric non-overlapping rings via extensive molecular dynamics simulations. The main idea is to study an assembly of model particles with anisotropic deformability using polymer rings. By tuning the degree of deformability of these model deformable particles, we study the dynamic and static properties of the assembly at different densities and temperatures. This deformable particle model might correspond to an assembly of epithelial cells or similar biologically soft bodies. In the limit at which the rings are very rigid with very little deformability, one expects to see the formation of a triangular lattice by the centres of these polymer rings. On the other hand, if one increases the deformability of these polymer rings, due to increased disorder, one observes glass-like dynamical behaviour even for identically sized polymer rings. We also show a transition from a crystalline state to a disordered glassy state driven solely by particle deformability. We observe non-trivial finite-size effects in the dynamics of these glass-forming ring polymers, not seen in usual molecular glass-formers.} 
\end{abstract}
\maketitle

%\end{titlepage}

\noindent{\bf \large Introduction:} \SK{Understanding glass-like dynamical response in various soft matter systems, especially in biological systems like a monolayer of confluent cells, for example, epithelial cells, intrigues the community for quite some time. A large body of research in active matter primarily focuses on the dynamical properties of collections of motile particles under different conditions to understand puzzling observations in biological systems. Observation of glass-like dynamical behaviour in confluent cells \cite{doi:10.1073/pnas.1010059108}, cell cytoplasm \cite{PhysRevResearch.1.032038}, leads to the study of active glasses \cite{Janssen_2019,paul2023dynamicalheterogeneityactiveglasses}, as often these systems are driven by non-thermal noises, like the conversion of ATP to ADP. Active glasses are known to show enhanced heterogeneity in their dynamics and, as a result, enhanced Mermim-Wagner effect \cite{paul2023dynamicalheterogeneityactiveglasses}. A substantial body of research on active glasses aims to improve the current understandings of glassy dynamics in biological systems. In all of these studies, it is often assumed that the constituent particles are isotropic; however, it is well-known that nearly all biological systems are soft and show anisotropic deformability \cite{annurev:/content/journals/10.1146/annurev.ph.49.030187.001141}. The question of whether the anisotropic deformability of these particles plays any role in their dynamics under various dynamical conditions remains poorly understood. Deformability in biological organelles is very important as life-threatening diseases like malaria\cite{Mohandas2012-je}, thalassemia, among others, are known to significantly reduce the degree of deformability or increase the stiffness of the Red Blood Cells (RBCs) \cite{10.1093/infdis/jiae490}, which leads to clogging in small passages like veins.}

\SK{Recently significant progress has been made in understanding the structure and dynamics of confluent cells using simple physics-based models like the vertex model \cite{PhysRevE.98.042418,10.7554/eLife.87966}. These studies have highlighted that many dynamical features like fluidization and jamming in these systems, which play important roles in various biological processes like wound healing, tumour growth, and stem cell differentiation, can be understood using simple physics-based models. Underlying many of these processes, there exists a transition in which cells become highly motile and lose confluency, known as the Epithelial to Mesenchymal transition (EMT) \cite{10.1172/JCI39104}. The vertex model, being a confluent model, is well-suited for studying the epithelial regime but fails to capture the essence of the EMT transition as non-confluency with the extracellular space plays a key role \cite{10.1172/JCI39104}.} 

\SK{Other popular models, like the Cellular Potts model, can accommodate the importance of these extracellular spaces in the modelling, but the surface fluctuations in the cell boundaries in the Cellular Potts model are shown to be unrealistic \cite{10.1371/journal.pone.0042852}. The Potts model, being a spin model, fails to explain the real-time dynamics of a realistic cellular system \cite{10.1371/journal.pone.0042852}. There is also a growing body of evidence that EMT is a continuous process rather than a transition \cite{Huang2022,10.1172/JCI39104}. Hence, an alternate model that can incorporate some of these caveats of other models, like tuning the cell-cell adhesion in the vertex model, will be very important in better understanding various important biological processes. The ring polymer model shows great potential for studying both confluent and non-confluent cells \cite{GRADA2017e11} and the transition between these two states. By introducing adhesion between the polymer rings, one can easily simulate a confluent system and investigate the dynamical features of both confluent and non-confluent layers of cells, which are known to affect the fluidity of the system \cite{ray2024roleintercellularadhesionmodulating,pasupalak2024epithelialtissuesbottomupcontact}. Furthermore, adding active noise to this model of polymer rings can mimic a close analogue of cells, where both the tendency to adhere to one another and their self-propelling energy significantly influence the structure and dynamics of these systems \cite{Schwarz_2013,FANG20202656}, and might play an important role in understanding the existence of leader cells in wound healing process \cite{doi:10.1098/rstb.2019.0391}. In other biological systems, structures such as plasmids (circular DNAs) found in bacterial cells \cite{doi:10.1128/mmbr.62.2.434-464.1998} and mitochondrial DNAs in some eukaryotic cells \cite{Guminska2022-gh} show similarities to this ring polymer model.} 

%The DNAs are by themselves not active. Hence for large enough rings, we expect the static factors, namely the bond-bond correlations to be similar to that of the experiments\cite{PhysRevLett.101.148103}, which indeed turns out to be the case from the initial results that we get from simulations.}

\SK{Moreover, there is also a growing interest in studying the mechanical and dynamical properties of embryonic tissues, as these properties play a crucial role in regulating embryo development \cite{10.1242/dev.190629}. Recent experiments have demonstrated that the viscosity of these tissues increases with the packing fraction until it reaches a critical point \cite{PETRIDOU20211914}. These tissues are characterised by their highly non-confluent and polydisperse nature. A recent soft deformable particle model has been utilised to explain the saturation of viscosity in relation to area fraction \cite{das2023freevolumetheoryexplains}. We believe that the ring polymer model with added polydispersity would be able to account for the experimental findings and the structural signatures observed in these systems.}

\SK{In this article, we utilise a ring polymer model similar to \cite{Gnan2019} to represent a deformable particle. An assembly of such ring polymers can be likened to various soft particle systems, such as soft colloids, emulsions, microgels and biological cells. Soft-colloids have been modelled through many polymeric models, like star and dendritic polymers. Ring-polymers have been used to simulate the soft-colloids, which can include asymmetry in the interactions among the rings \cite{doi:10.1073/pnas.2122051119,10.1063/1.4866644}. These ring-polymer models have the advantage of possible mapping to cells due to their closed topology. Soft colloids are known to exhibit a range of intriguing phases like glassy states \cite{doi:10.1021/mz5006662}, and re-entrant behaviour \cite{Gnan2019}. Jamming transitions have been extensively studied for regular point particles. But tuning the softness of these particles can introduce interesting behaviours and phases, such as cluster crystals, where the rings nearly overlap \cite{PhysRevE.97.052614}. A few similar studies in the athermal limit have used deformable polygon models. A crucial parameter in these systems is the ratio of the softness to temperature, known as the "softness parameter". Particulate models, such as Hertzian particles, fail to capture a crucial aspect of the system: soft particles shrink and even interpenetrate at higher densities, which cannot be accounted for by merely tuning interaction strengths. We demonstrate here that, at certain densities, glassy dynamics can be observed solely due to the deformability of the rings, which introduces structural frustration leading to dynamical slowdown.} 

%Adding points on dimensionality and jamming.
%Dimensionality

\SK{The study of glass transition in this model with packing fraction has been explored \cite{Gnan2019}. However, the behavior concerning temperature remains largely unexamined, even though temperature is a critical factor in determining the "softness parameter". In this work, we extensively investigate this aspect of the model. The dependence of relaxation times on the temperature of soft colloids is known to be linked to the fragility of the system, which we calculate. Recently, in the context of the vertex model, researchers have studied the relaxation profile in relation to activity, demonstrating a very good correlation with the mode-coupling theory (MCT) framework, as reported in \cite{PhysRevE.111.054416}. We conducted a similar analysis concerning temperature and found similar results to hold although at lower temperatures, one observes some deviations. We observe a growing dynamic and static characteristic in these systems as temperatures decrease, consistent with the behavior of supercooled liquids prior to glass transition \cite{doi:10.1073/pnas.0811082106}.}

%Similar models of ring polymers( statics and dynamics)( \cite{10.1063/1.3587137,10.1063/1.3587138,PhysRevLett.73.1263,PhysRevLett.99.198102}) have been studied extensively where in the dynamics the effect of the ring length is compared with that of the line polymer for their effect in diffusion constants, relaxation times, stress relaxation, viscosity and entanglement length while the radius of gyration structure factor dependence on the number of monomers and the shape of a semiflexifle ring polymer of studies in the statics. But the glassy regime still remains unexplored in this model.

\SK{The finite size effects in this soft particle model  especially at lower temperatures are puzzling, as we observe a re-entrant behavior in the mean squared displacement (MSD) relative to system size. This effect appears to intensify as the temperature decreases. To our knowledge, this phenomenon has not been observed in any equilibrium systems with such clarity. Interestingly, there is no significant difference in the static markers across different system sizes at these lower temperatures. This effect also seems to be independent of the density at which the system is simulated and appears to be a universal feature at lower temperatures. At these temperatures and for densities close to the Hexatic-order peak, the Van Hove correlation function appears to develop shoulders at distances approximately equal to $R_g$, and this effect persists (see the supplementary material). The re-entrant behavior also becomes more pronounced when we increase the stiffness of the rings. A conclusive explanation for this effect remains elusive and requires further investigation. At higher temperatures, the shoulders disappear at these lower densities, and we revert to a behavior characteristic of supercooled glass, where the Van Hove functions display Gaussian characteristics \cite{PhysRevLett.99.060604}.}

%\PB{Softness as stated can be tuned by tuning the "softness parameter". We did comparisons of the static factors at higher stiffness and lower temparatures. By decreasing the softness in a high enough density, we expect to get an emerging crystalline behaviour. By means of hexatic order parameter, indeed we see that with stiffness we get an emergent crystalline behaviour and also mark a sweet spot in density where the crystalline behaviour is most pronounced.}

\SK{The rest of the paper is organised as follows. We first describe the model and simulation details, then present the results. We discuss the glass-like dynamical behaviour in the system and then study the reentrant glass transition. We then focus on the finite-size effects and compute both dynamic and static correlation lengths in the supercooled regime. Subsequently, we study the amorphisation transition in this model by systematically tuning the degree of deformability of each ring. Finally, we conclude with a discussion of the significance, future prospects and implications of these results.}

%It is possible to cite multiple references at the same time %\citep{collins2011b,collins2016,lunn2007a,lunn2007b,ross2006,shannon1948}.

\section{Model and Simulation Details}
\label{mm}
We study the system of non-overlapping polymer rings in two dimensions, where the interaction potentials are as follows, 
%\begin{align*}
\begin{equation}
U(r) = 4\epsilon\left[\left(\frac{\sigma}{r}\right)^{12}-\left(\frac{\sigma}{r}\right)^{6}\right]+ \epsilon, r\le r_c 
\end{equation}
which acts between all the beads for $r<r_c=2^{\frac{1}{6}}\sigma$

\begin{figure*}[!hptb]
  \raggedright
  \includegraphics[width=1.0\textwidth]{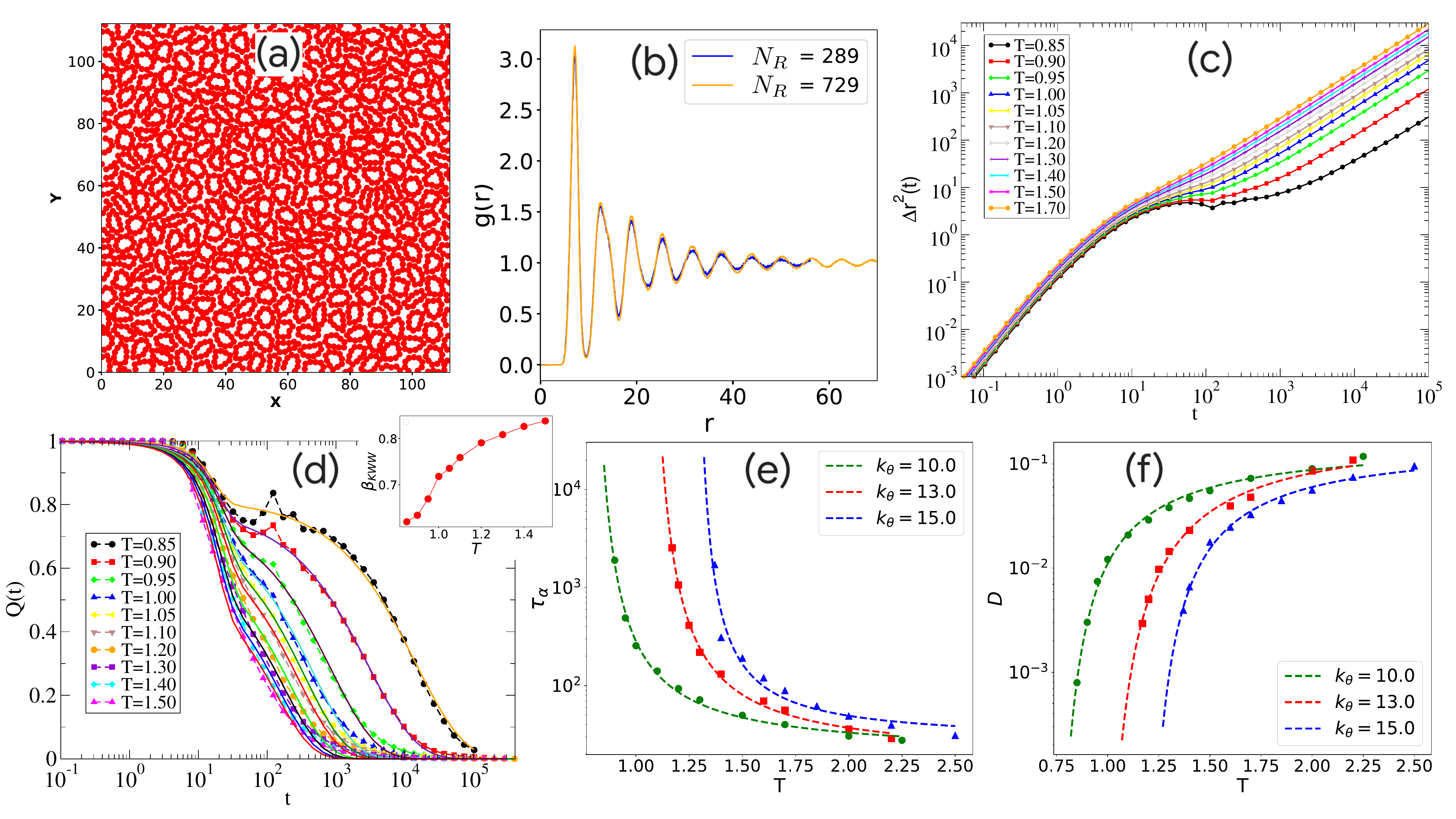}
  \caption{{\bf Glassy Dynamics of Ring-polymer Assembly.} \small(a) A typical configuration of the rings in a dense deformed state. One sees that the centre of mass (CoM) of these rings forms a disordered structure. (b) The radial distribution of the CoMs for two different system sizes ($N_R =289$ and $N_R=729$), which are practically indistinguishable. (c) The MSD plots for various temperatures for $k_{\theta}=10.0$ are plotted. The appearance of a clear plateau at lower temperatures indicates the onset of landscape-dominated glassy dynamics and caging effects. (d) The self-overlap functions, $Q(t)$, with stretched exponential fits show the hallmark two-step relaxation process typical of glass-forming liquids. The inset shows the temperature dependence of $\beta_{kww}$, the stretching exponent. (e) The VFT fits of $\alpha$-relaxation times at three different $k_{\theta}s$. (f) The corresponding diffusivity fits for the same $k_{\theta}s$ at a monomer density of $\rho = 0.23$.}
  \label{basic}
\end{figure*}
We use the FENE potential to model for finite extensibility\cite{Smrek2020}.
\begin{equation}
U_{FENE}(r) = -0.5k_F{R_F}^2{\sigma_m^2}ln\left(1-\left(\frac{r}{R_F\sigma_m}\right)^2\right)
\end{equation}
The angular forces that act at each angle of the ring (i.e between two consecutive bonds of a ring)
\begin{equation}
 U_{angle}(\theta) = k_{\theta}(1-cos(\theta-\theta_0))
\end{equation}
\SK{For the simulations, we set the parameters as follows: $\sigma_m = 1.8$, $\epsilon = 1.0$, $k_F = 15.0$, and $\theta_{0} = 144^\circ$, which is the equilibrium angle for the initialized decagon pattern. We kept the number of monomers in each ring fixed at 10. While maintaining a constant $k_F$, we varied $k_\theta$ from $10.0$ to $100.0$. Our initialization protocol involved placing the rings on a square lattice and then running an NPT simulation to achieve the desired density (adjusting the length slightly to reach this density). Following this, we allowed the system to equilibrate for $6.5 \times 10^6$ steps—significantly longer than any relaxation time of the system. Both NPT and NVT simulations were conducted using the Nose-Hoover method \cite{20919}. Each ring had a radius of $\approx3.2$, with monomers positioned at equal angles. It is important to note that the effective temperature of each ring polymer, considered as a single entity, is the same as the simulated temperature for the system with $10$ monomers. The effective system remains in equilibrium, as the effective temperature does not change when transitioning from NVT to NVE simulations. Unless otherwise stated, all simulations were conducted with $289$ ring polymers, using LAMMPS\cite{LAMMPS}. For the smallest system sizes, we did 64 ensembles and 150 time origin averaging, and gradually changed the number of ensembles to 4 for the largest system sizes, with the same number of time origin averaging.}

\section{Results and Discussions}
\noindent{\bf \large Supercooled Behaviour of the Ring Polymer Assembly:}
\SK{The supercooled liquid preceding the glass transition is known to exhibit dramatic increases in its relaxation times; an associated growth of a static length scale ($\xi_s$) has been attributed to this increase, but a clear picture has yet to emerge \cite{doi:10.1073/pnas.0811082106}. Other dynamical quantities that are typically associated with supercooled behaviour are stretched exponential decay of two point density-density correlation function which is often approximated using the self-overlap functions, $Q(t)$, increasing peak in the dynamic susceptibility ($\chi_4(t)$) along with a growing dynamical correlation lengths, $\xi_d$, peaks in the non-Gaussianity parameter ($\alpha_2(t)$) at intermediate times, the deviation from usual Gaussian form of the van Hove correlation function at larger displacement leading to the near universal exponential tail, etc \cite{10.1063/5.0166404}. We characterise glass-like dynamical features in this ring-polymer model using some of these well-known dynamical and static quantifiers. Definitions of the dynamical quantities $Q(t)$, $\chi_4(t)$, $\alpha_2(t)$ and van Hove function are given in the Supplementary Materials (SM).} 

\SK{In Fig.\ref{basic}(a), we show a typical configuration of an assembly of these ring polymers. One can clearly see that the centre of mass (CoM) of each particle forms a disordered structure even though the ring-polymers are identical in nature. The main reason behind this disorder structure is solely due to their degree of deformability, as we show later that one can transition to a crystalline phase by systematically decreasing the degree of deformability via the strength of the angle-dependent potential term, $k_\theta$ (specifically around the hexatic order peak density). It is interesting to see how deformability alone can induce enough frustration in the system so as to completely destabilise the crystalline ground state (triangular lattice in this case). To further quantify the structural aspect of the assembly, we compute the radial distribution function, $g(r)$, as shown in Fig.\ref{basic}(b). We see the absence of any structural order in $g(r)$. Results for two different system sizes, one with a number of rings $NOR = 289$ and another with $729$ rule out any finite-size effects in the structure.}

\SK{Now to characterise the dynamics, we first compute the mean squared displacement (MSD), $\Delta r^2(t)$ (see SM for definition) as a function of time. We clearly see all the important characteristics of glassy dynamics in the MSD. At higher temperatures, the system shows usual liquid-like behaviour with MSD showing ballistic behaviour at short times and diffusive at longer timescales. On the other hand, at lower temperatures, the effects of supercooling and slow dynamics begin to emerge through the formation of a plateau in the MSD curve as shown in Fig.\ref{basic}(c). This hallmark caging behaviour in MSD clearly indicates that the system is exhibiting glass-like dynamical behaviour. We further compute the self-overlap correlation function $Q(t)$ as shown in Fig.\ref{basic}(d). Once again, one can clearly see a plateau in the $Q(t)$ vs $t$ curve, indicating the well-known two-step stretch-exponential relaxation process observed in typical glass-forming liquids. }

\SK{We fitted $Q(t)$ data using the fitting form
\begin{equation}
 Q(t)=a e^{-(t/{\tau_s})^2}+(1-a) e^{-(t/{\tau_{\alpha}})^{\beta_{kww}}},
\end{equation}
to obtain the degree of stretching in the relaxation function and the associated stretching exponent, $\beta_{kww}$. $\beta_{kww}$ is known as the Kohlrausch–Williams–Watts exponent of the stretched exponential, which decreases as we decrease temperature, and it goes from $0.9$ at high temperatures to nearly $0.6$ at the lowest studied temperature, as shown in the inset of Fig.\ref{basic}(d). From the fitting function, we can obtain $\tau_s$, which is associated with the short-time decay of the overlap function to the plateau and $\tau_{\alpha}$, associated with the long-time eventual decay of the correlation function, the alpha-relaxation time. The monomer density of the regime of the glassy behaviour is fixed at 0.23.}
\begin{figure}[!b]
  \raggedleft
  \includegraphics[width=1.00\columnwidth]{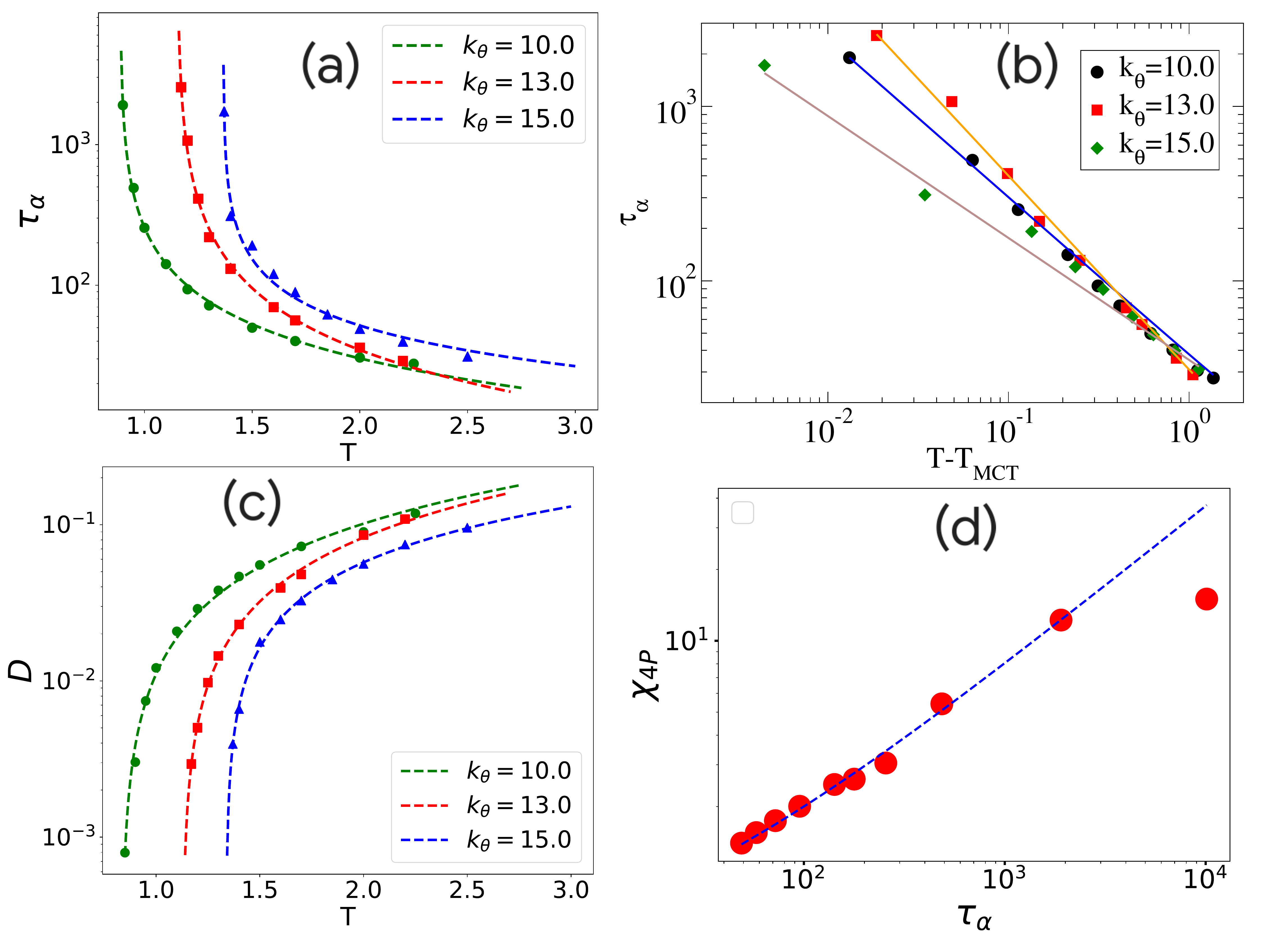}
  \caption{{\bf Correspondence with Mode Coupling Theory Predictions.} (a) The fits for power law divergence of the relaxation times as predicted in MCT, (b) The same power law fits to $\tau_\alpha$ vs $T-T_{MCT}$. It is interesting to see that $\tau_\alpha$ obeys MCT predictions very well in the entire temperature window for all three values of $k_\theta$. (c) The fits to diffusivity ($D$) vs temperature follow the predicted MCT form. The fits are found to be very good. (d) The power-law fit to the $\chi_4(t)$ peaks vs. $\tau_\alpha$, as predicted by MCT ($N_R=289$). It becomes poor at lower temperatures, indicating a similar deviation seen in many molecular glass-forming liquids.}
  \label{MCT}
\end{figure}

\begin{figure*}[!htpb]
  %\raggedright
  \includegraphics[width=1.00\textwidth]{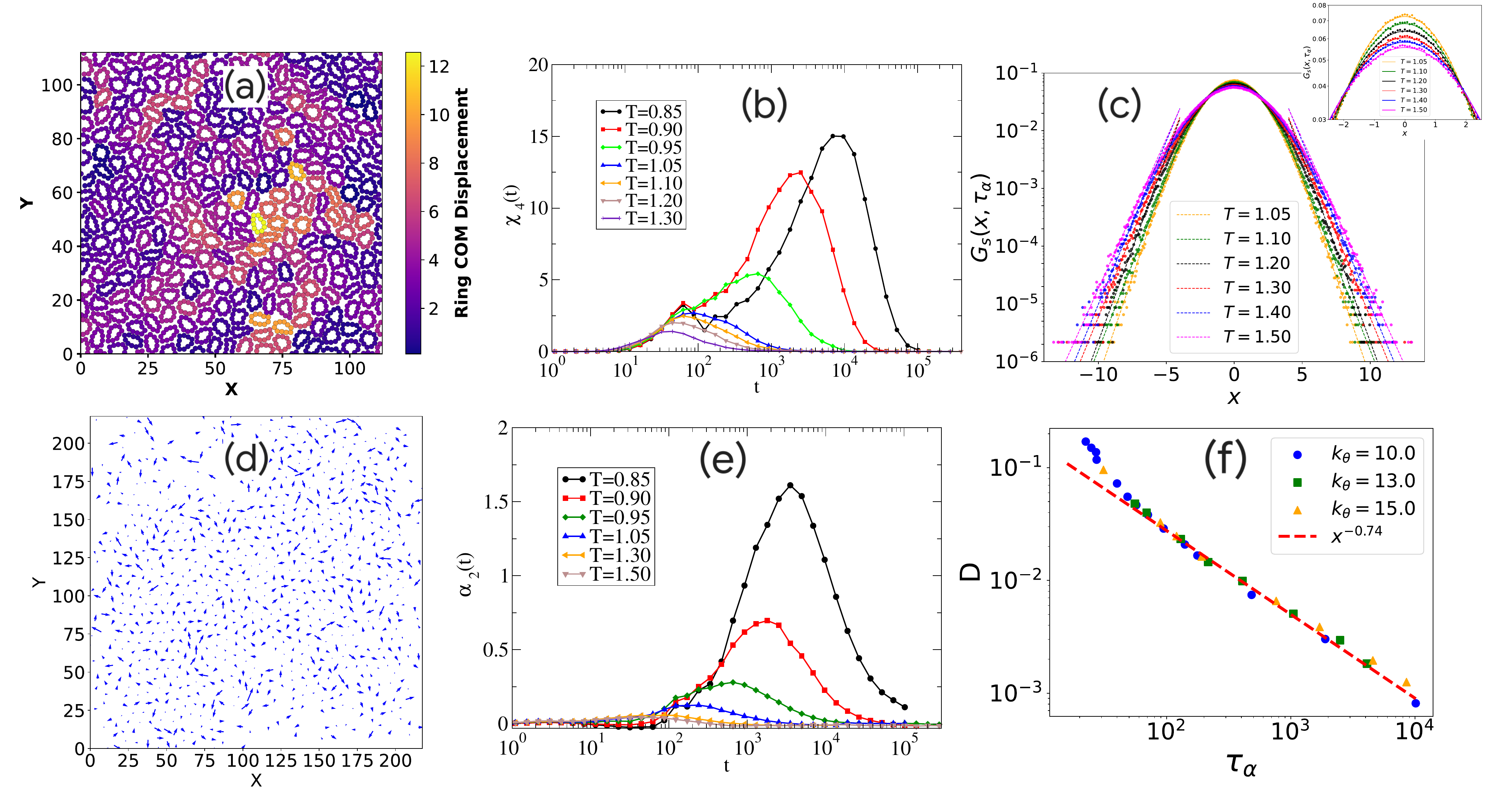}
  \caption{{\bf Dynamical Heterogeneity.} (a) A typical configuration of rings where the rings are coloured according to the distance moved in a time interval to demonstrate dynamic heterogeneity. (b) The increase in the peak height of the four-point dynamic susceptibility ($\chi_4(t)$) with decrease in temperature, (c) The van-hove function for different temperatures (Fits in the main plot are exponential fits and the plot in the inset are the Gaussian fits to the van-hove function), (d) The displacement field of the COMs for a specific time period which clearly shows the displacement fields are quite heterogeneous ($N_R=1089$). (e) The non-Gaussianity parameter ($\alpha_2(t)$) also shows the same trend as that of $\chi_4(t)$. (f) The breakdown of the Stokes-Einstein relation for three different $k_{\theta}$s. The $\kappa$ exponent in , $D\propto\tau_{\alpha}^{-\kappa}$, is $\approx 0.74$ with very slight change among $k_{\theta}$s (The data is obtained for $N_R=289$).}
  \label{DH}
\end{figure*}
\SK{In Fig.\ref{basic}(e), we showed the obtained alpha-relaxation time, $\tau_\alpha$ as a function of temperature. It shows the usual characteristic of glass-forming liquids of a dramatic increase in relaxation time with decreasing temperature. We show results for three different strengths of deformability of the particles characterised by $k_\theta$. We then fit the structural alpha-relaxation time $\tau_{\alpha}$ to the empirical VFT (Vogel-Fulcher-Tamman) equation for the studied temperature range. The VFT equation reads as  
\begin{equation}
 \tau_\alpha = \tau_0 \exp \left [ \frac{1}{K_{VFT}({T}/{T_{VFT}}-1)}\right],
\end{equation}
with $K_{VFT}$ being the kinetic fragility, which characterises how fast the relaxation time grows with decreasing temperature, and $T_{VFT}$ is the extrapolated divergence temperature. One sees that with increasing $k_\theta$, the divergence temperature increases, as well as a mild increase in the kinetic fragility $K_{VFT}$. The fragility parameter $K_{VFT}=0.0306,0.0345,0.0442$ for $k_{\theta}=10.0,13.0,15.0$ respectively. Similarly, the diffusivity ($D$) obtained from MSD as $\Delta r^2 (t) = 4Dt$ from large time data, can also be fitted to a similar VFT relation as 
\begin{equation}
    D = D_0\exp\left[-\frac{1}{K^{'}_{VFT}(T/T_{VFT}-1)}\right]
\end{equation} 
as shown in Fig.\ref{basic}(f).}

\SK{In a recent work \cite{PhysRevE.111.054416}, it was argued that the relaxation behaviour of the vertex model, which shows similarity with the dynamics of epithelial cells, can be well described by Mode Coupling Theory (MCT) predictions. They also validated their results using experiments. To test this in our ring-polymer model, we fitted the relaxation-time and diffusion-coefficient data using MCT predictions. According to the predictions of MCT, the relaxation times are known to diverge as a power law with temperature as 
\begin{equation}
\tau_{\alpha} \sim (T-T_{MCT})^{-\gamma_1}, \,\,\, D \sim (T-T_{MCT})^{\gamma_2}
\end{equation}
where $T_{MCT}$ is the MCT transition temperature and $\gamma_1$ and $\gamma_2$ are the power-law exponents. The $\gamma_1$ exponents are 0.938, 1.138, 0.702; and $\gamma_2$ exponents are 1.137, 1.111, 0.896; and $T_{MCT}$ are 0.861, 1.1325, 1.352 for $k_\theta=10.0,13.0,15.0$ respectively (The $T_{MCT}$ values were very close for the diffusivity(D) and relaxation times($\tau_\alpha$) fits and we averaged the two $T_{MCT}$ values to report these values). Fig.\ref{MCT}(a) and (c) show the power-law fit of the relaxation time and diffusivity data over the studied temperature. The observed fits are quite good for all $k_\theta$. After extracting the MCT divergence temperatures, we showed the same data in a log-log plot, and it is interesting to see that relaxation time data obeys power-law behaviour in the entire temperature window, in agreement with the observation made in \cite{PhysRevE.111.054416} as shown in Fig.\ref{MCT}(b) ( For this we used the $T_{MCT}$ obtained from temperature divergence only). There are some deviation at lower temperatures. Also, according to the MCT prediction, the four-point susceptibility peak, $\chi_{4}^{P}$, should show a power-law dependence on the relaxation times as $\chi_{4}^{P} \sim \tau_{\alpha}^\eta$, which agrees with our simulation data as shown in Fig. \ref{MCT}(d), except for the last temperature.}

\vskip +0.1in
\noindent{\bf \large Dynamical Heterogeneity:} \SK{In the supercooled-liquid phase, it is known that there is spatial heterogeneity in the displacements of particles, where some regions are dynamically very mobile while other regions remain almost immobile. This effect in the context of epithelial cells has been previously observed specifically in the context of wound healing as of recently \cite{doi:10.1098/rstb.2019.0391}, where the presence of dynamic heterogeneity is shown to influences the leader-follower dynamics. A similar signature of strong dynamical heterogeneity is observed in this model. To quantify this effect and its associated correlation volume, we first compute the four-point dynamic susceptibility, $\chi_4(t)$ (see definition in the SM), which shows a clear peak at a timescale comparable to the alpha-relaxation time, and the peak height increases with decreasing temperature. In Fig.\ref{DH}(a), we show a typical configuration of our ring-polymer model at a monomer density of $\phi = 0.23$ with colour coding given on the side. High mobility is marked as yellow, with blue being small mobility. The cluster of yellowish regions in various parts of the sample clearly suggests the existence of strong dynamical heterogeneity in the system, in close analogy with typical glass-forming liquids. Fig.\ref{DH}(b) shows the four-point susceptibility, $\chi_4(t)$, at different temperatures, and one clearly sees a strong growth of the peak in line with the growth of the dynamical heterogeneity in the system with further supercooling. Fig.\ref{DH}(c) shows another universal characteristic of the glassy dynamics, namely the non-Gaussian displacement distribution as quantified by the van Hove correlation function, $ G_s (x,\tau_\alpha)$, computed at the peak position of $\chi_4(t)$. One clearly sees a Gaussian behaviour at the central region ($  x\simeq 0$) with universal exponential behaviour at large $x$. Typical displacement field pattern is shown in Fig.\ref{DH}(d). To quantify the degree of non-Gaussianity in the dynamics, we computed the well-known non-Gaussian parameter, $\alpha_2(t)$, which is plotted in Fig.\ref{DH}(e). It also shows a strong peak at timescale, which is comparable to $\alpha$-relaxation time, and shows again strong growth with decreasing temperature. Similar studies with large ring-polymers have been shown in \cite{10.1063/5.0160097}, where the rings had $50$ monomers, but our focus is to model these rings as deformable particles, instead of studying the dynamics of the rings itself as done in the \cite{Gnan2019}.}

\SK{Now, we focus our attention on the Stokes-Einstein (SE) violation in our model system. The SE relation relates relaxation time to diffusivity in an equilibrium system as 
\begin{equation}
\frac{k_BT}{D}\propto\tau_{\alpha}.   
\end{equation} 
It is well known in the literature \cite{10.1063/1.4792356} that the SE relation is violated in supercooled liquids, and there are ample evidences that suggest that the primary reason behind SE breakdown is solely due to the presence of dynamic heterogeneity in the system \cite{PhysRevE.108.L022601}. One intuitive way to understand this breakdown is as follows: the relaxation time is dominated by the slow-moving particles, whereas the diffusivity is controlled by the fast-moving particles. As soon as there is a dynamical decoupling of these two quantities due to the presence of dynamic heterogeneity, one immediately expects to see the SE breakdown. It was also noted in recent works \cite{PhysRevLett.119.056001, PhysRevE.108.L022601} that one can associate a length scale above which the SE relation will be valid, and below that length scale, the SE will be violated at any given temperature. This length scale is shown to be the dynamic heterogeneity length scale. Often in the literature, one studies a slightly modified version of the SE relation, known as the fractional SE relation, defined as 
\begin{equation}
D\propto\tau_\alpha^{-\kappa},    
\end{equation}
with $\kappa < 1$ being the fraction SE exponent.  In a wide variety of supercooled liquids,  the relation is found to be true with $\kappa<1$ in both 3D and 2D for lower temperatures \cite{10.1063/1.4792356}. In our model system, we get $\kappa \approx 0.74$ with a slight change across stiffness, which corroborates well with the presence of dynamic heterogeneity in the system as shown in Fig. \ref{DH}(f). The power-law is found to be very good over at least two orders of magnitude of variation in $D$ and $\tau_\alpha$.}

\begin{figure*}
  \raggedleft
  \includegraphics[width=1.00\textwidth]{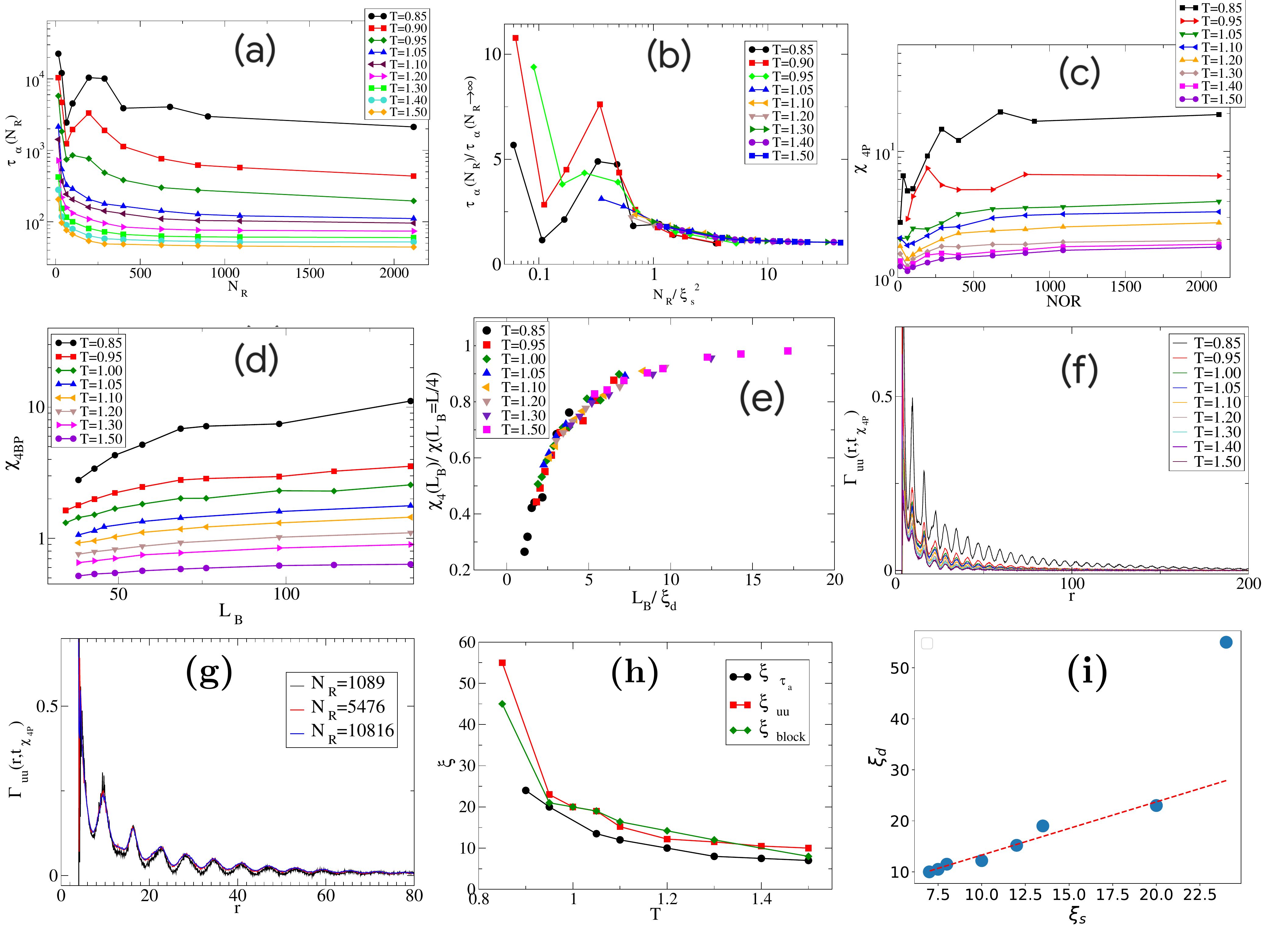}
  \caption{{\bf Non-trivial Finite Size Effects.} (a) The $\alpha$-relaxation times ($\tau_\alpha$) variation with system size for different temperatures. Notice the non-monotonic behaviour at an intermediate system size at lower temperatures (see text for detailed discussion). (b) The collapse of the relaxation times data for all the system sizes using a growing static length scale, $\xi_s$, as elaborated in the text, (c) The system size dependence of the $\chi_{4P}$s for different temperatures, which shows a clear sign of increase with system size. One observes some noise at intermediate system sizes, very similar to the system sizes where one sees non-monotonic behaviour of $\tau_\alpha$. (d) The $\chi_4$ peaks obtained by the block analysis method are discussed in the main text. Notice the improvement to the signal-to-noise ratio in this method. (e) The collapse of the scaled $\chi_4(L_B)$ with $L_B/\xi_d$. The data collapse is quite good, suggesting that the obtained length scale will have lower uncertainty. (f) The displacement-displacement correlation function, $\Gamma_{uu}(r,t_{\chi_{4P}})$, is plotted as a function of $r$ for all the studied temperatures. One clearly sees that the correlation increases with decreasing temperature, as the decay of the correlation function becomes slower. (f) The finite size effect of $\Gamma_{uu}(r,t_{\chi_{4P}})$ at $T=0.95$. There are no finite-size effects in $\Gamma_{uu}(r,t_{\chi_{4P}})$. (h) The growth of the dynamic and static length scales obtained from the different methods is plotted as a function of temperature. One clearly sees a strong growth of these length scales with decreasing temperature. (i) Cross plot of $\xi_d$ and $\xi_s$ to highlight the concomitant growth of these length scales in accordance with MCT-like dynamical behaviour with breakdown at lower temperatures.}
  \label{FSS}
\end{figure*}

\vskip +0.1in
\noindent{\bf \large Finite Size Effects:}
\SK{We now systematically study the finite-size effects in our measured quantities. Given that dynamical heterogeneity grows rapidly with decreasing temperature, one expects the underlying correlation length scales to grow, leading to strong finite-size effects in various dynamical quantities, including the four-point susceptibility. We first study the finite-size effects in the $\alpha$-relaxation time. As in usual glass-forming liquids, it has already been reported that the relaxation time shows strong finite-size effects due to the growth of the underlying static length scale, $\xi_s$ \cite{doi:10.1073/pnas.0811082106, Karmakar2014, PhysRevE.86.061502}. In Fig.\ref{FSS}(a), we have plotted $\tau_\alpha$ as a function of number of rings, $N_R$ for various temperatures. At higher temperatures, one observes $\tau_\alpha$ to decrease with increasing system size in complete agreement with results reported in the literature \cite{doi:10.1073/pnas.0811082106, PhysRevE.86.061502}. Interestingly, with increasing supercooling, one observes non-monotonic behaviour in the system-size dependence. For temperatures $T\le 0.95$, one observes that $\tau_\alpha$ first decreases for very small system sizes, and then it starts to increase at intermediate system sizes, and then eventually decreases for very large system sizes. It shows a clear peak at an intermediate system size, with the peak position shifting to larger system sizes as temperature decreases. This clearly indicates a growth of a static correlation length in the systems. Although we do not have a good microscopic understanding of this non-monotonic dependence, we tried to collapse the large system size data using a static length scale as shown in Fig.\ref{FSS}(b). The large-system-size data collapse very nicely onto a master curve, and the corresponding length scale is shown in Fig.\ref{FSS}(f). We will discuss this matter in greater detail later.
}

\SK{Next, we study the finite size effects on four-point dynamic susceptibility, $\chi_4(t)$. In Fig.\ref{FSS}(c), we showed how the peak height of $\chi_4(t)$ (referred to as $\chi_4^P$) varies with system size. $\chi_4^P$ is found to increase with system size, eventually saturating at a large system size limit. One observes some noisy behaviour at intermediate system size precisely at those system sizes where one sees non-monotonic behaviour in $\tau_\alpha$. As $\chi_4(t)$ computed in the canonical ensemble lacks many important fluctuations like temperature, number of particles, and shape fluctuations for this model, we compute $\chi_4(t)$ using the block analysis method \cite{PhysRevLett.119.205502, PhysRevLett.105.015701}, in which one computes the four-point susceptibility in a subsystem embedded in a bigger system. This can be thought of as an effective grand canonical ensemble. We show $\chi_4^P$ variation as a function of subsystem size (referred to as $L_B$) in Fig.\ref{FSS}(d). One sees a nice variation in $\chi_4^P$ with increasing $L_B$. The noise in this calculation is expected to be low due to improved averaging, as discussed in \cite{PhysRevLett.119.205502}. We then attempted a finite-size scaling collapse of these data sets by appropriately choosing the large $L_B$ value of $\chi_4^P$ and a dynamical length scale, $\xi_d$. The data collapse is observed to be quite good as shown in Fig.\ref{FSS}(e). So the estimated length scale can be considered reliable. We then use this length scale to collapse the data obtained from conventional canonical ensemble simulations. The data collapse is presented in the SM. The variation of $\xi_d$ with temperature is plotted in Fig.\ref{FSS}(h) labelled as $\xi_{block}$.}
\begin{figure*}[!htp]
  %\raggedleft
  \includegraphics[width=0.99\textwidth]{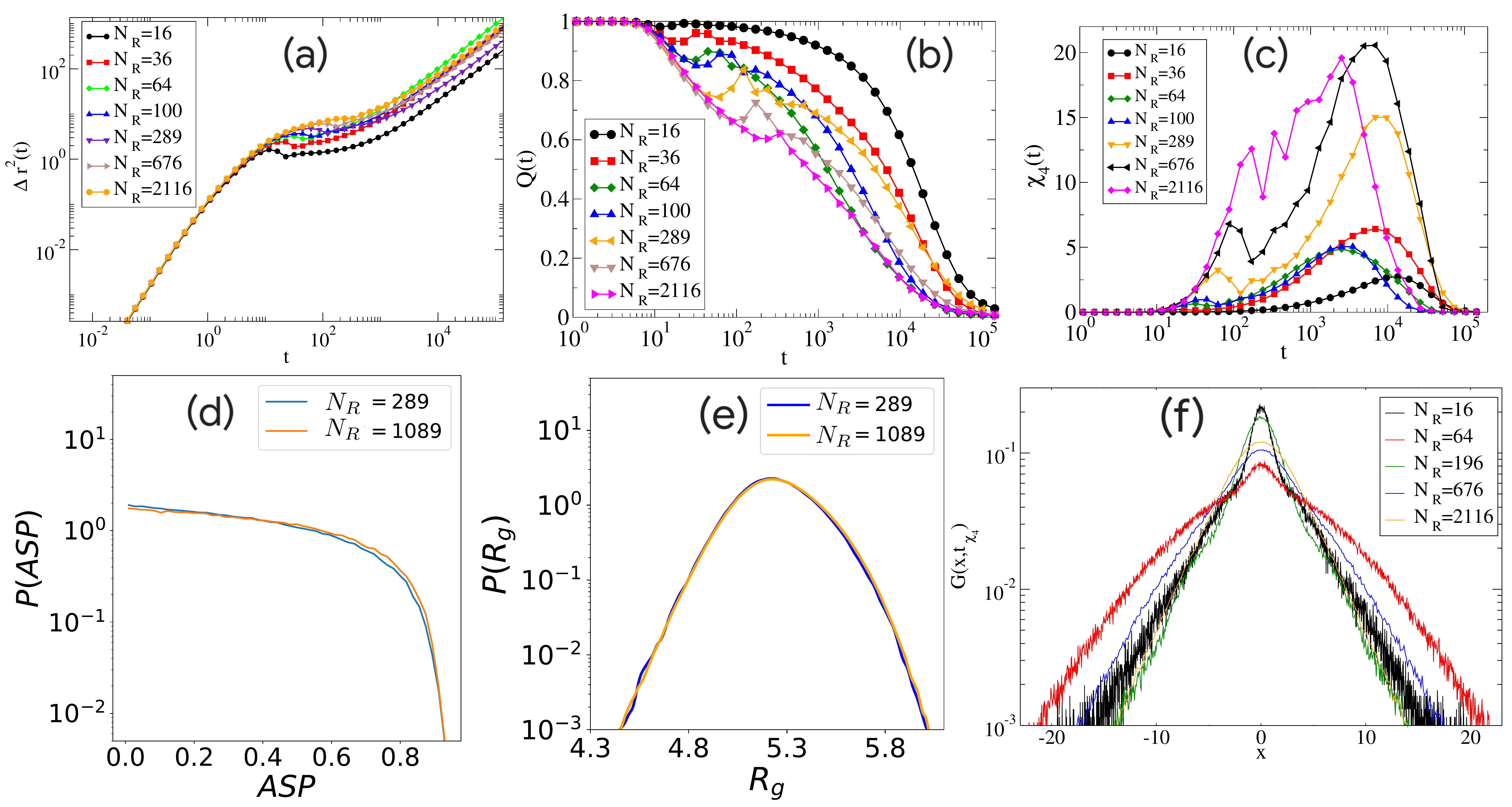}
  \caption{{\bf Reentrant Dynamical Crossover with System Size.} (a) Shows the MSD for various system sizes, indicating a crossover in the dynamics from slower to faster to slower. (b) highlights the same crossover in the self-overlap correlation function. (c) Similar results for $\chi_4(t)$ with system size. All these analyses are done at $T=0.85$. At large system sizes, the $\chi_4(t)$ develops multiple peaks at short timescales and becomes noisy because of phonons. (d), (e) show the ASP and $R_g$ distribution for large and small system size, which show no difference, (f) The van Hove functions obtained at $t_{\chi_{4P}}$, the peak position of $\chi_4(t)$.}
  \label{reensysmsd}
\end{figure*}

\SK{To reaffirm the growth of this dynamical length scale, we computed the displacement-displacement correlation function, $\Gamma_{uu}(r,t_{\chi_{4P}})$ as shown in Fig.\ref{FSS}(f). We can find the dynamic length-scale from the spatial correlation of the COM displacements (see SM for details), which basically gives us the measure of how much heterogeneity there is in the system with respect to the CoM displacements of the rings, that is, the difference in the mobility of different regions in the system. Indeed, as expected in normal glass formers, we observe a length scale that increases with decreasing temperature, similar to the results in \cite{PhysRevResearch.2.022067,D2SM00727D}.
The plot clearly shows that $\Gamma_{uu}(r,t_{\chi_{4P}})$ systematically increases as the correlation function decays to larger values of $r$. Fig.\ref{FSS}(g) shows how the correlation function behaves for different system sizes. There are no appreciable finite-size effects, so an estimate of the correlation length scale from $\Gamma_{uu}(r,t_{\chi_{4P}})$ will not be influenced by finite-size effects. The correlation length is computed by simply integrating the $\Gamma_{uu}(r,t_{\chi_{4P}})$ to remove any fitting-related error. The estimated correlation length is found to be in good agreement with the length scale computed from block analysis as shown in Fig.\ref{FSS}(h) (referred to as $\xi_{uu}$). We have also checked the dynamic length scale from the spatial correlation of mobility ($S_4(\vec{q},t)$), and the length scale matches the order of magnitude. In Fig.\ref{FSS}(i), we show a direct comparison of $\xi_s$ and $\xi_d$. We find that the static correlation length grows in the same manner as the dynamic correlation length as found in a recent study in the active Vertex model \cite{PhysRevE.111.054416} and some models with Medium-range-crystalline order (MRCO) (see \cite{PhysRevLett.121.085703}). We see some deviation at the lowest temperature studied.}

\SK{To gain a better understanding of the non-monotonic finite-size effects on both relaxation time and four-point susceptibility, we closely examine the mean squared displacement (MSD) of the system at $T = 0.85$ for various system sizes, as shown in Fig. \ref{reensysmsd}(a). Typically, in glass-forming liquids, the MSD increases with system size; however, in this model at low temperatures, we observe a re-entrant behavior in the MSD. It initially grows slowly for intermediate system sizes before accelerating at larger sizes. The overlap correlation function $Q(t)$ exhibits the same behavior, as illustrated in Fig. \ref{reensysmsd}(b). In panel (c), we present the evolution of $\chi_4(t)$, which shows a systematic increase in the peak. However, for larger system sizes, the peak begins to oscillate at shorter timescales. Notably, a short-time peak emerges for larger system sizes, indicating the presence of strong phononic activity, similar to what is observed in typical glass-forming liquids in two dimensions (2D) \cite{PhysRevLett.116.085701}. This short-time peak in $\chi_4(t)$ has been attributed to phononic fluctuations, which are particularly pronounced in 2D systems due to the Mermin-Wagner-Hohenberg (MWH) theorem \cite{dey2024enhancedlongwavelengthmerminwagner}. It is particularly interesting that this model, involving deformable particles, also exhibits a strong signature of the MWH theorem.} 

\begin{figure*}
  \raggedleft
  \includegraphics[width=1.00\textwidth]{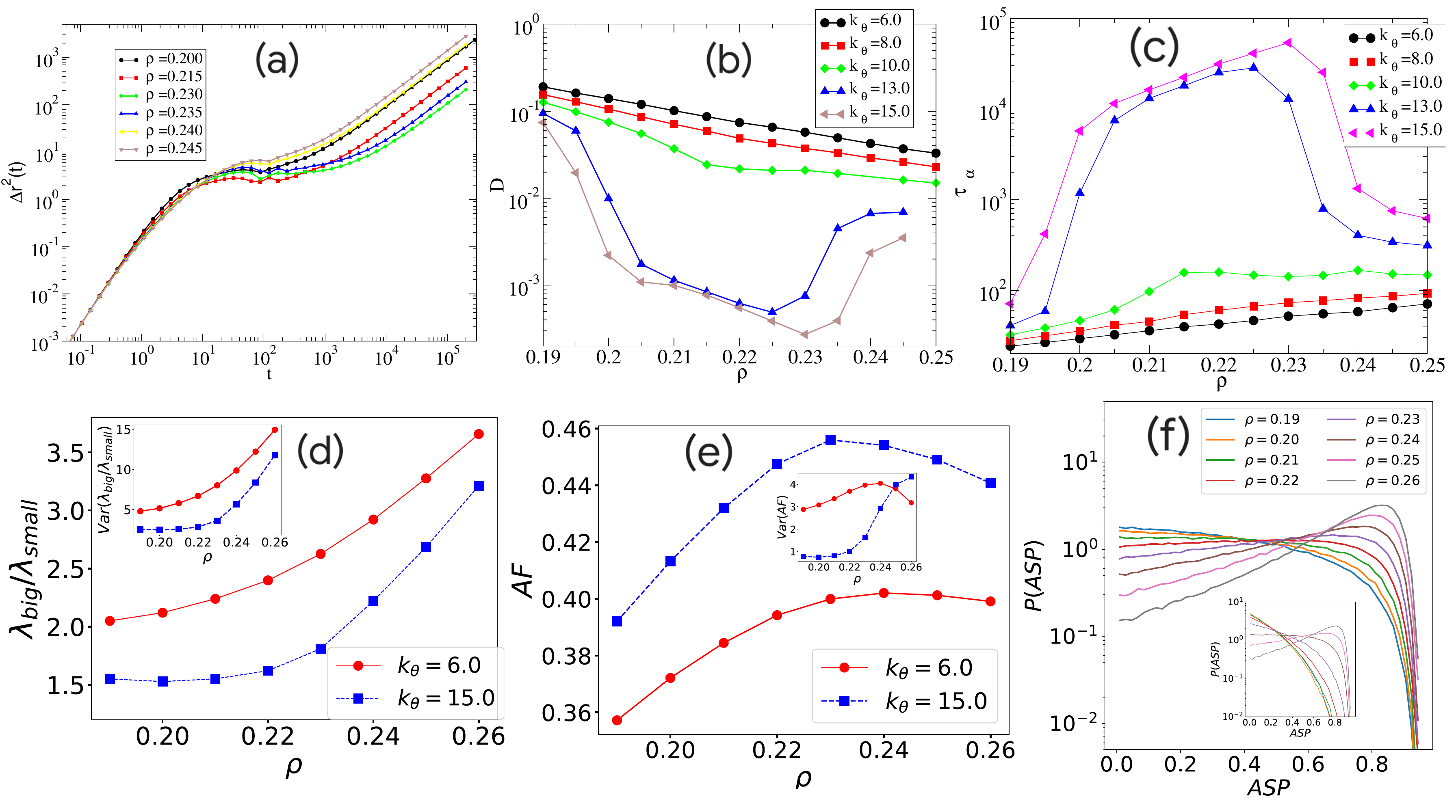}
  \caption{{\bf Reentrant Dynamical Crossover with changing Density.} (a) The re-entrant behaviour in MSD at $k_{\theta}=15.0$, (b) The plots of diffusivity vs. density at $T=1.00$, (c) The relaxation times at different densities, (d) The ratio of the largest to the smallest eigen vector for the highest and lowest $k_{\theta}$s, inset shows standard deviation of the same. This indicates that as density increases, the rings become progressively more aspherical, which is the primary reason for the reentrant behaviour observed in the system. (e) The area fraction vs. density for the highest and lowest $k_{\theta}$s, with the inset showing the standard deviation of the same. The greater the standard deviation of both quantities, the greater the polydispersity in the system. (f) The asphericity parameter distribution for the highest and lowest $k_{\theta}$. The inset is for $k_{\theta}=15.0$ and the main plot is for $k_{\theta}=6.0$, where the change in the distribution is very gradual.}
  \label{reen}
\end{figure*}
\SK{We also characterise other structural features, such as the asphericity parameter (ASP) and Radius of gyration ($R_g$) \cite{10.1063/5.0160097} of these rings and their respective distributions. To define these two quantities, we first define a two-dimensional matrix 
\begin{equation}
S_{pq}=\frac{1}{N}\sum_{i=1}^{N}(r_p^i-R_{p}^i)(r_q^i-R_{q}^i),
\end{equation}
for each ring where $p$ and $q$ are the cartesian coordinates in two dimensions, and they take two values $x$ and $y$. $N$ is the number of monomers ($N = 10$) in this case. $R_{p}^i$ is the centre of mass position, respectively.
The asphericity parameter is then defined as :
\begin{equation}
    ASP  =  \frac{(\lambda_1^2-\lambda_2^2)^2}{(\lambda_1^2+\lambda_2^2)^2} \\
  =  1-\left(\frac{2c}{b^2-2c}\right)^2
\end{equation}
where, $\lambda_1$ and $\lambda_2$ are the eigenvalues of the shape-matrix and $b=-(S_{xx}+S_{yy})$, $c=(S_{xx}S_{yy}-S_{xy}S_{yx})$. As the rings become more circular, the ASP approaches 0.
The radius of gyration is defined as
\begin{equation}
    R_g  = (\lambda_1^2+\lambda_2^2)^{\frac{1}{2}} \\
     = (b^2-2c)^{\frac{1}{2}}
\end{equation}
The bigger the rings are, the larger the radius of gyration.
}

\SK{In Fig.\ref{reensysmsd}(d) and (e), we show the distribution of ASP and $R_g$, respectively, for two different system sizes. It is clear that there is not much difference in the Asphericity parameter in the system with increasing system size, nor in the radius of gyration of the rings. Thus, some of these structural aspects, which otherwise play a crucial role in the formation of glassy states, are not responsible for the puzzling system size dependence in relaxation time. Finally, we show the van Hove correlation functions in Fig.\ref{reensysmsd}(f) across the studied system sizes, and one also sees the same non-monotonic behaviour in the particle displacement. Thus, observed non-monotonicity is very robust but does not seem to be directly related to the structural aspect of these constituent rings.}

\vskip +0.1in
\noindent{\bf \large Dependence of Relaxation Dynamics on the Stiffness and Re-entrant Behaviour:}
\SK{We now explore the changes in the dynamics as a function of density or packing fraction for various degrees of stiffness. In \cite{Gnan2019}, it has already been reported that this model system shows a re-entrant behaviour with increasing density or packing fraction. In usual systems with isotropic interactions, one typically observes an increase in relaxation time with increasing density, with a close connection to the Jamming transition in the athermal limit. This model, in contrast, shows a different behaviour with increasing density solely due to deformability. As one increases density at a fixed $k_\theta$, the system slows down very rapidly and then at much larger density it fluidises once again, as can be seen from the dependence of MSD on density in Fig.\ref{reen}(a). It shows slower diffusion at intermediate density range and then becomes faster at densities close to $\rho = 0.23$ and above, as demonstrated in Fig.\ref{reen}(b).}

\SK{Relaxation time also echoes the same behaviour. $\tau_\alpha$ first increases sharply in the density range $\rho = 0.19$ to $\rho = 0.205$ and then shows a mild increase in the density range $\rho \in 0.21 - 0.23$, followed by a sharp drop fro $\rho \ge 0.23$. This reentrant behaviour is very strong for $k_\theta = 15.0$ and $13.0$ and becomes a bit weak for $k_\theta \le 10.0$. To link this re-entrant behaviour with the changes in the structural properties, we once again look at the ASP and $R_G$. In Fig.\ref{reen}(d), we show the ratio of large and small eigenvalues of the $S_{pq}$ matrix. It is clear that $\lambda_{big}/\lambda_{small}$ shows a strong increase for $\rho\ge0.23$, indicating a sharp change in the shapes of the polymer rings. This is also highlighted in the area fraction (AF) vs density plot in Fig.\ref{reen}(e). Area fraction in this refers to the area covered by the polymer rings as their asphericity changes with increasing density.} 

\SK{For $k_\theta = 15.0$, one sees a strong rise in AF until density $\rho = 0.23$, and then it starts to decrease after that, clearly linking a sharp decrease in relaxation time for $\rho\ge0.24$. The inset shows the variation of AF with $k_\theta = 15.0$, with the sharpest change occurring at increasing density. This also highlights that the system will exhibit strong dynamic heterogeneity, as discussed in the first part of the article. Finally, in Fig.\ref{reen}(f), the distribution of ASP, $P(ASP)$, is plotted for all the studied densities, and it also shows that with increasing density, ASP increases steadily, with the distribution peak shifting towards large values. Thus, the re-entrant behaviour can be understood as the consequence of shape change from spherical to ellipsoidal \cite{PhysRevLett.110.188301}. Although we are able to understand the re-entrant behaviour as a function of density, the re-entrant behaviour observed for system size still remains a puzzle. Recently, a decrease in viscosity has been linked to a decrease in the area fractions of these systems. Our results can be mapped onto those results \cite{10.7554/eLife.87966}.}

%To quantify this we computed the ratio of the largest and smallest eigen-values in these systems across the re-entrant for the smallest $k_{\theta}=6.0$ and largest $k_{\theta}=15.0$. These is a flat portion across the densities for the highest $k_{\theta}$s initially, that is the rings resist change of shapes which makes them more crystalline and increase the relaxation times and then there is a sharp change in the ratio which causes the re-entrant. Similarly, the area fractions decrease much more( more free space) for higher $k_{\theta}$s than lower $k_{\theta}$s, which can also account for the re-entrant. The average increase can be accounted for by considering the increase in the density which is the case for normal liquids. Recently, a decrease in viscosity has been linked to decrease in the area fractions of these systems. Our results can be mapped onto those results( \cite{10.7554/eLife.87966}). The diffusivities show the opposite trend to that of the relaxation times, which is plausible. We also plotted the distribution of the asphericity of these systems. A similar picture arises where the change in asphericity distribution with density for $k_{\theta}=15.0$ is much more steeper than $k_{\theta}=6.0$ where the change is more gradual.(see fig. \ref{reen})

 \begin{figure*}
%  \raggedleft
  \includegraphics[width=0.995\textwidth]{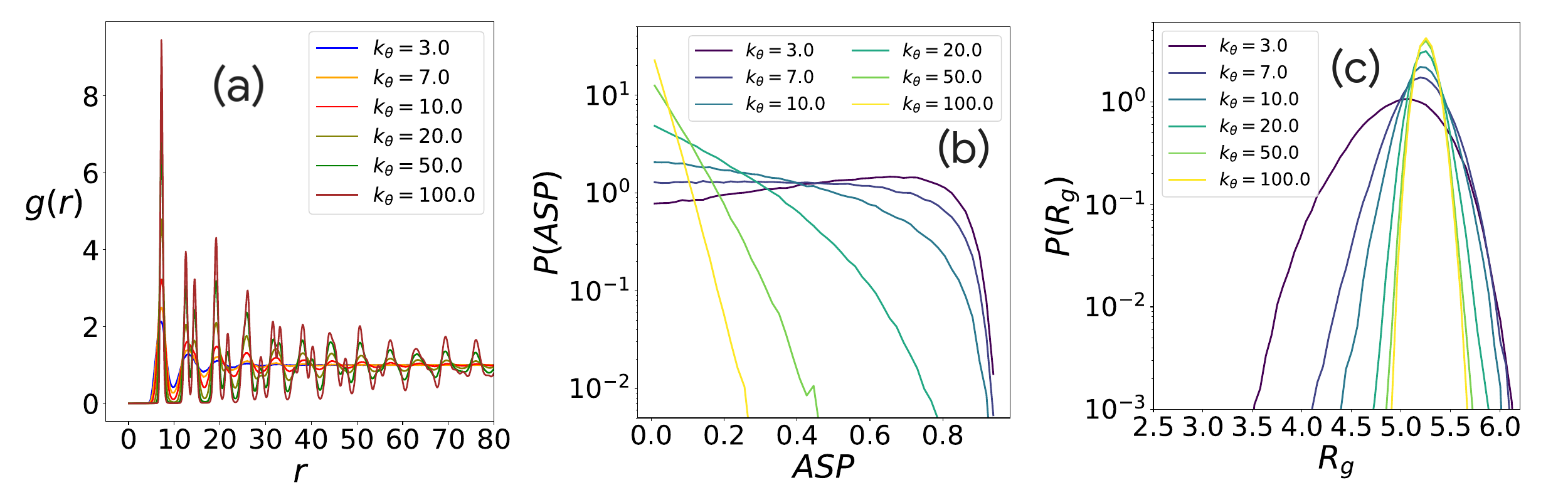}
  \caption{{\bf Amorphisation Transition.} (a),(b),(c) are the radial distribution functions, asphericity parameter distribution and radius of gyration distribution for various $k_{\theta}s$ at $T=1.00$, $\rho=0.22$. These static structural indicators clearly show the amorphisation transition as a function of increasing deformability.}
  \label{Statfactors}
\end{figure*}
\vskip +0.1in
\noindent{\bf \large Crystal to Glass Transition with Deformability:}
\SK{The effect of changes in phase properties of the systems is examined as the degree of deformability is varied systematically by adjusting $k_\theta $. At large $ k_\theta $, the rings can be considered as rigid circles, and it is anticipated that the system will crystallize into a triangular lattice. This suggests a deformability-induced amorphization transition as a function of $k_\theta$. At smaller values of $k_\theta$, glass-like dynamical behavior is expected, while at larger $k_\theta$, a crystalline phase is observed at lower temperatures. In Fig.\ref{Statfactors}(a), the radial distribution function, $g(r)$, is presented for various $k_\theta$ values in the range of $3.0$ to $100.0$. Sharp peaks in $g(r)$ are noted with increasing $k_\theta$. Figures \ref{Statfactors}(b) and (c) depict the distribution of the aspect ratio (ASP) and $R_G$ for the studied $k_\theta$ values. As expected, with increasing $k_\theta$, the rings display increased circularity, resulting in a systematic decrease in both ASP and $R_G$, accompanied by a narrowing of the respective distributions.}

\begin{figure*}[!htpb]
  \includegraphics[width=0.995\textwidth]{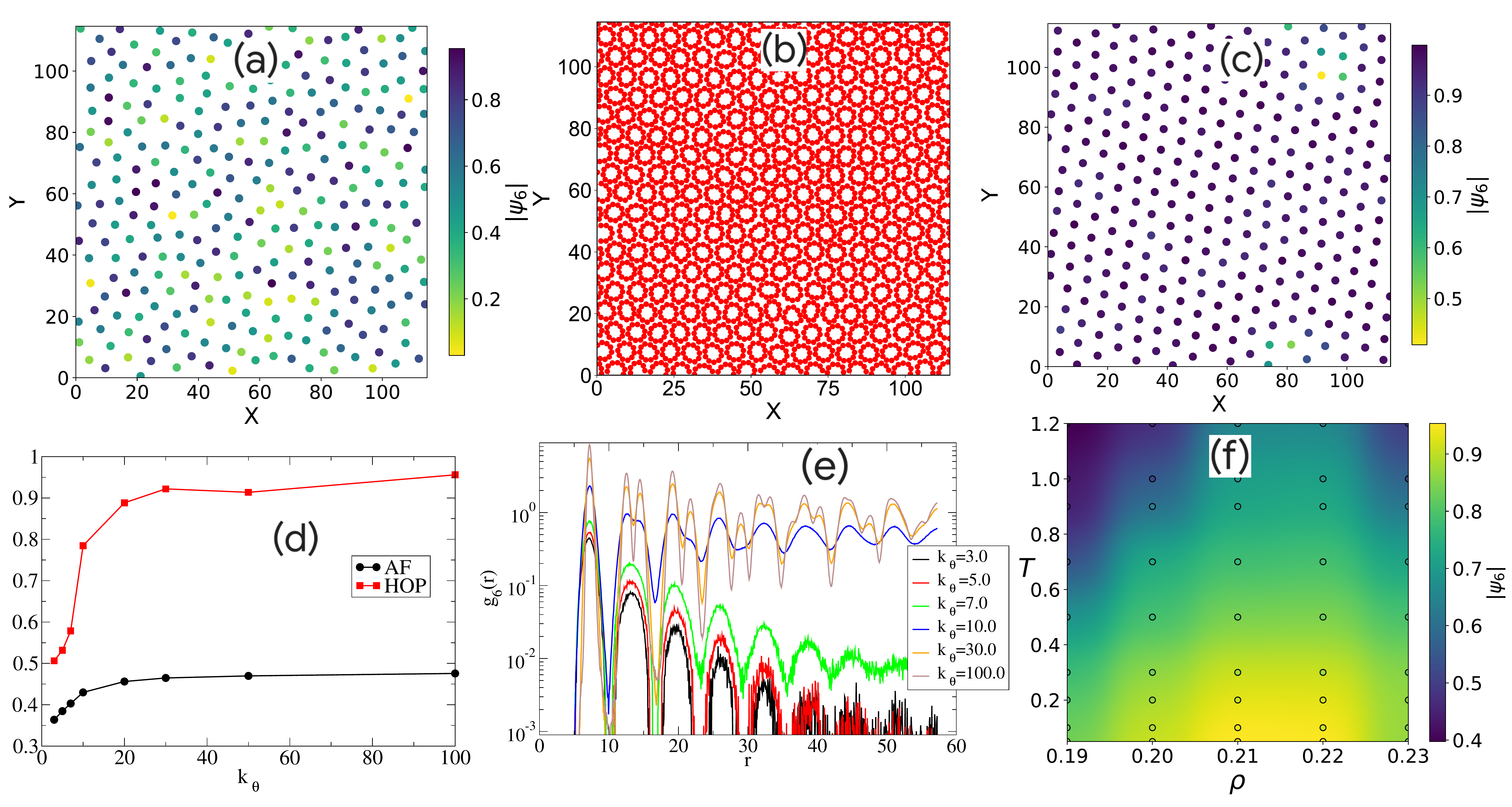}
  \caption{{\bf Phase Diagram.} (a) Typical configuration showing the CoMs of rings with their Hexatic values in a disordered state shown by points for $k_{\theta}=5.0$, (b) An actual sample configuration of the ring at $\rho=0.22$ for $k_{\theta}=100.0$, showing the formation of triangular lattice. (c) Plot of a sample configuration of CoMs for $k_{\theta}=100.0$ ($N_R=289$), (d) The Area fraction and HOP magnitude for $\rho=0.22$ versus $k_{\theta}$, indicating a sharp crossover from amorphous to crystalline state with increasing $k_\theta$. (e) The spatial Hexatic correlation function, $g_6(r)$ for various $k_{\theta}$, shows that hexatic correlation increases with $k_{\theta}$, and a clear crossover in structure at around $k_\theta \ge 7.0$. (f) Heat map of magnitude of $\psi_6$ for $k_{\theta}=10.0$ in the $\rho - T$ plane.}
  \label{HOPCOM}
\end{figure*}

\SK{The amorphisation transition is analyzed in greater detail. The Hexatic order parameter is computed to identify the critical $k_\theta$ at which the system transitions from the glassy state to the crystalline state. Figure \ref{HOPCOM}(a) presents a typical configuration of the center of mass (CoM) of a ring-polymer assembly with $k_\theta = 5.0$. The CoMs of each ring exhibit a disordered structure. CoMs of each particle are colored according to the Hexatic order parameter value, $|\psi_6|$. In two dimensions, the Hexatic order parameter is defined as
\begin{equation}
\psi_6^i=\frac{1}{N_b}\sum_{j=1}^{N_b}e^{\imath 6\theta_{ij}}, 
\end{equation}
where $N_b$ is the number of nearest neighbors of particle $i$, and $\theta_{ij}$ is the angle between particles $i$ and $j$ relative to the x-axis. The hexatic order parameter clearly indicates that even the local environment surrounding each ring is disordered. In Fig. \ref{HOPCOM}(b), we display the ring configuration for a system with $k_\theta = 100.0$. This indicates that the rings are very rigid and less deformable, allowing them to be treated as approximately spherical particles. Therefore, it is expected that at sufficiently high density, they will form a triangular lattice in two dimensions, which we observe in the results. Panel \ref{HOPCOM}(c) of the same figure illustrates the centers of mass of each ring arranged in a triangular lattice configuration. 
}

\SK{To determine whether the observed amorphization phenomena, as a function of deformability, represent a gradual transition or a sharp crossover, we calculate the average hexatic order parameter, defined as $|\psi_6| = \frac{1}{N}\sum_{i=1}^N |\psi_6^i|$, where N is the number of frames times number of rings, as illustrated in Fig. \ref{HOPCOM}(d). The results show that the hexatic order parameter increases sharply around $k_\theta \simeq 16.0$. Additionally, the area fraction (AF) is plotted in the same figure panel. It indicates that the area fraction gradually increases as the system begins to exhibit local ordering.}

\SK{Next, we study the spatial correlation between the local Hexatic order parameter as it can clearly indicate the emergence of long-range order as the system crosses over to the crystalline state with increasing $k_\theta$. In Fig.\ref{HOPCOM}(e), we show the Hexatic-order correlation function, $g_6(r)$ \cite{dey2024enhancedlongwavelengthmerminwagner} defined as 
\begin{equation}
g_6(r)=\frac{1}{\rho N_R}\langle\sum_{i\neq j}\psi_6(\vec{r_i})\psi_6^*(\vec{r_j})\delta(r-|\vec{r_i}-\vec{r_j}|)\rangle
\end{equation}
for different $k_{\theta}$. We can see that for smaller $k_\theta \le 10.0$, the correlation function decays exponentially, but for $k_\theta\ge 30.0$, the correlation does not decay, indicating the emergence of hexatic order in the system. A recent study \cite{kumar2025particledeformabilitystabilizeshexatic}, showed that deformability stabilises the hexatic order in these 2D ring-polymer systems at large packing fraction. Our results are in complete agreement with this recent study.}
 
\SK{We construct a phase diagram using the global average Hexatic order parameter in the $T-\rho$ plane. In Figure \ref{HOPCOM}(f), we illustrate the phase diagram, where symbols represent simulated data points, and the heat map colour code indicates the Hexatic order parameter of the entire system. Yellow corresponds to higher values of the Hexatic order parameter. We observe that at lower temperatures, there is a re-entrant transition from the disordered phase for $\rho \leq 0.19$ to the crystalline phase in the intermediate density range of $0.19 < \rho \leq 0.24$, followed by a return to the disordered phase at larger densities of $\rho > 0.24$, for $k_\theta = 10$. Another interesting point is that increasing the number of monomers in the rings shifts the Hexatic peak to lower density due to stronger excluded-volume effects, but these large rings start to show nematic ordering at higher densities\cite{10.1063/5.0160097} which is absent in small rings.}

%\subsection{The behaviour of rings at extremely low density :}
%We expect that the rings will behave as particles moving in an ideal gas at low densities. Here, we consider the rings having a mass of 10 units and calculate the pressure using the kinetic energy of the COMs with the ideal gas law( i.e from the dynamics) and the temperature is the temperature at which we simulated the system.(see Fig. \ref{Gas}) 
%\begin{figure}[!htpb]
%  \raggedright
%  \includegraphics[width=1.00\columnwidth]{Fig6.pdf}
%  \caption{
%{\small (a) Gaussian distribution of $v_{x,COM}$, pre-exponents of which are 5.015622,6.248597,8.339957,12.525599,24.928339 for temperatures T=1.0,0.8,0.6,0.4,0.2 (For the ideal case, $\frac{m}{2T}=5.00,6.25,8.33,12.50,25.00$ which are quite close to that of the fits) (b) Pressure vs. temperature of rings polymers considered as a single entity at $\rho=0.01$, which is linear as expected for an ideal gas}}
%  \label{Gas}
%\end{figure}

\vskip +0.1in
\noindent{\bf \large Conclusions:}
\SK{To conclude, we demonstrated that the ring-polymer model in two dimensions is a highly versatile framework to study various systems whose constituents have deformability as an additional degree of freedom. There are a large number of systems in which the particles possess a certain degree of deformability, which plays an important role in determining their dynamics and the variety of complex phases. This model will be a very good minimal model to understand non-confluent cell monolayers or soft colloidal materials, with the possibility to also design a core-shell colloidal particle model by putting an additional stiff polymer ring inside a floppy ring. This, we believe, might give us a minimal model system to study various puzzling behaviours, including two-step yielding found in many core-shell soft colloidal particles in experiments\cite{doi:10.1021/acs.langmuir.6b03586}. Our study confirms that this simple model can exhibit a plethora of phases across various parameter ranges, including a glassy phase in a certain density and deformability window.}

\SK{In the glass-like dynamical regime, this model system shows all the hallmark features of glass-forming liquids like stretched exponential relaxation and appearance of a clear plateau in the mean squared displacement at lower temperatures. The temperature dependence of the relaxation time is well described by the empirical Vogel-Fulcher-Tamman (VFT) relation. Interestingly, this system in the studied range shows dynamical features that are well described by Mode Coupling Theory (MCT) predictions, especially power-law-like divergence in the relaxation time with temperature for various degrees of deformability. Moreover, it shows strong dynamical heterogeneity as evidenced by non-Gaussian behaviour in the van Hove function and the non-Gaussianity parameter. The system also exhibits the well-known Stokes-Einstein (SE) Breakdown, which is directly linked to the underlying dynamical heterogeneity. The system is found to obey a fractional SE relation with an exponent close to $-0.74$, which is much smaller than those observed in many molecular glass-forming liquids.}

\SK{Another key characteristic of this system is its non-trivial finite size effects observed in relaxation time, especially for lower temperatures. One observes the relaxation time to first decrease with increasing system size and then increase at intermediate system size, eventually saturating to a smaller value at large system size, giving rise to a broad peak at a characteristic system size at different temperatures. This non-trivial system size dependence is observed in the kinetically constrained model (KCM), where the distance to the MCT can be tuned by changing a parameter in the model, as discussed in \cite{PhysRevE.71.026128}. This KCM model showed an increase in relaxation time with increasing system size, akin to MCT-like critical dynamical behaviour. This study then argues that one can observe a non-monotonic system-size dependence in the relaxation time if, at some larger system size, activated dynamics becomes important. Thus, one expects the non-monotonic finite-size effects to be more pronounced near the MCT transition temperatures. We indeed find that the system at higher temperatures shows usual monotonic behaviour, but as one approaches the MCT transition temperature, one starts to see non-monotonic finite size effects in complete agreement with the arguments in  \cite{PhysRevE.71.026128}. In the same work, it was shown that the soft sphere model interacting via a Harmonic repulsive interaction shows some (albeit weak) signature of this non-monotonic behaviour. Our model appears to be an ideal candidate, showing the same crossover in dynamics as a function of system size; thus, we expect the same behaviour to be visible in vertex model simulations. It will be interesting to see if one observes similar dynamical features in cell-monolayer experiments.}

\SK{Another highlighting feature of this model is the concomitant growth of the static and dynamic length scales. It is found that they are proportional to each other in the studied temperature range, in close agreement with recent results from the vertex model simulation \cite{PhysRevE.111.054416}. The model also supports the existence of many phases and transitions, especially the amorphisation transition with a changing degree of deformability. It also shows a re-entrant dynamical behaviour at an intermediate density range where the system crosses from fluid phase through supercooled liquid phase and back to fluid phase due to changes in its asphericity parameter. A similar re-entrant behaviour is found in colloidal experiments with spherical and ellipsoidal particle mixture \cite{PhysRevLett.110.188301}. The re-entrant in that case was also induced by changing the number of ellipsoidal particles and thus due to the asphericity parameter. In future, it will be interesting to study various dynamical fluctuations in the systems, especially the existence of long wavelength fluctuations due to the famous Mermin-Wagner-Hohenberg (MWH) theorem. We already have some intriguing evidence that MWH are important in this system, as $\chi_4(t)$ shows growth of a peak at short timescales.}

\SK{Finally, it will be interesting to understand the primary differences in the dynamics of ring polymers when transitioning from two to three dimensions. This change might come from the topological constraints, such as no-threading or looping in two dimensions, as well as excluded-volume effects. A suitable deformable particle model in three dimensions (3D) will be very interesting to develop to understand whether some of these observations can be translated to 3D. A deformable particle model in 3D will be of significant interest, as it might serve as a useful minimal model to study biological soft tissues, which are not constrained in 2D, or the rheological properties of deformable particles, especially their flow through channels.}

\vskip +0.1in
\noindent{\bf \large Acknowledgements:}
We would like to thank Santu Nath, Surajit Chakraborty, Antik Bhattacharya and Roni Chatterjee for many useful discussions. We would also acknowledge funding from intramural funds at TIFR Hyderabad, provided by the Department of Atomic Energy (DAE) under Project Identification No. RTI 4007. SK acknowledges Swarna Jayanti Fellowship grants DST/SJF/PSA01/2018-19 and SB/SFJ/2019- 20/05 from the Science and Engineering Research Board (SERB) and Department of Science and Technology (DST). Most computations are performed using the HPC clusters procured through Swarna Jayanti Fellowship grants DST/SJF/PSA01/2018-19 and SB/SFJ/2019- 20/05. SK would like to acknowledge the research support from the MATRICES Grant MTR/2023/000079, funded by SERB. SK also acknowledges travel support from the JSPS invitational fellowship.

\bibliographystyle{apsrev4-2}
\bibliography{main}

\end{document}

%\preprint{APS/123-QED}
%\begin{titlepage}
\title{Supplementary material :}% Force line breaks with \\

\author{Padmanabha Bose}

\author{Smarajit Karmakar}
\affiliation{Tata Institute of Fundamental Research Hyderabad}
%\end{titlepage}
%\affiliation{%


%\affiliation{
%\affiliation{
%%\affiliation{%


\maketitle

%\tableofcontents

\section{Definitions of observable quantities :}

Various quantities have been calculated in this paper. We present the detailed definitions here\cite{doi:10.1073/pnas.0811082106,D2SM00727D}.
\subsection{Mean squared displacement( MSD(t)) :}
\begin{equation*}
    MSD(t)=\frac{1}{N_R}\sum_i^{N_R}|\vec{r}_{i,COM}(t)-\vec{r}_{i,COM}(0)|^2 
\end{equation*}
where $\vec{r}_{i,COM}(t)$ is the position of the ith COM of a ring in open coordinates, and $N_R$=number of rings. When there is a flattening of the MSD(t) vs t curve, we say that it is reminiscent of caging behaviour.
\subsection{Two-point Self-overlap-function :}
\begin{equation*}
    Q(t)=\frac{1}{N_R}\sum_i^{N_R}w(|\vec{r}_{i,COM}(t)-\vec{r}_{i,COM}(0)|-a) 
\end{equation*}
where w is the window function and a is the cutoff ($a^2=6.5$). Stretched exponential type-function at large times is the hall-mark of glassy dynamics.

\subsection{Four-point suseptibility function :}
\begin{equation*}
    \chi_4(t)=N_R.(\langle Q^2(t)\rangle-\langle Q(t)\rangle^2)
\end{equation*}
which captures the fluctuations in the overlap function. Peaks in $\chi_4(t)$ is reminiscent of a large rearrangement at a specific time at the length scale probed using a. Higher peaks are tantamount to rearrangement of bigger chunks of particles displacing at one time.

\subsection{Cage-relative mean-squared displacement :}

\begin{equation*}
    MSD_{CR}(t)=\frac{1}{N_R}\sum_i^{N_R}|\vec{r}_{i,COM}^{'}(t)-\vec{r}_{i,COM}^{'}(0)|^2 
\end{equation*}
where $\vec{r}_{i,COM}^{'}(t)=\vec{r}_i(t)- \frac{1}{N_{nn}} \sum_j^{N_{nn}}(\vec{r}_i(t)-\vec{r}_j(0))$, where $N_{nn}$ is defined as the number of neighbours defined as the number of particles within the first dip in the peak of g(r). This form is known to suppress the log(L) divergence in the MSD pleateau, with system size(L). 

\subsection{Non-gaussianity parameter :}

\begin{equation*}
    \alpha_2(t)=\frac{\langle\Delta r_{COM}^4\rangle}{2\langle\Delta r_{COM}^2\rangle^2}-1 
\end{equation*}
 which gives a measure of the deviation from gaussian dynamics. At the cage breaking event the deviation is maximum and hence gives a peak in this parameter.
\subsection{Van Hove function :}
\begin{equation*}
    G(x,t)=\frac{1}{N_R}\sum_i^{N_R}\delta(x-|\vec{r}_{i,COM}(t)-\vec{r}_{i,COM}(0)|)
\end{equation*}
where delta is the 1 when the argument is 0. Here we use a softer version with some wiggle room for the delta which only becomes one when it has an argument close to 0 determined, by the specific binning. 

\subsection{Spatial correlation of COM displacements :}
The spatial correlation used to quantify the dynamic heterogeneity length scale is defined as (Here $\rho_{COM}$ is the COM number density)
\begin{equation*}
    \Gamma(r,t)=\frac{g^{uu}(r,\Delta t)}{g(r)}-1   
\end{equation*}
where,
\begin{equation*}
    g^{uu}(r,\Delta t)=\frac{\langle\sum_{i,j}^{N_R}u_i(0,\Delta t)u_j(0,\Delta t)\delta(r-|\vec{r}_{ij}(0)|)\rangle}{4\pi \Delta r .r^2.N_R.\rho_{COM}.\langle u(\Delta t)\rangle^2} 
\end{equation*}
where,
\begin{equation*}
\begin{split}
    u_i(t,\Delta t)&=|\vec r_i(t+\Delta t)-\vec r_i(t)|\\ \langle u(\Delta t)\rangle&=\langle\frac{1}{N_R}\sum_i^{N_R}u_i(t,\Delta t)\rangle
\end{split}
\end{equation*}
and 
\begin{equation*}
    g(r)=\frac{\langle\delta(r-|\vec{r}_{ij}(0)|)\rangle}{4\pi \Delta r .r^2.N_R.\rho_{COM}} 
\end{equation*}
\subsection{Block analysis :}
If we divide a large system into small sub-systems, then the averaging is better due to the grand-canonical nature of the subsystems. We can define the self-overlap function for one block as below :
\begin{equation*}
 Q(L_B,t)= \frac{1}{n_i}\sum_{j=1}^{n_i} [w(r_{j,COM}(t)-r_{j,COM}(0))],
\end{equation*}
where $L_B= (N_R/N_B)^{1/3}$ and, $N_B$ is the number of subsystems referred henceforth as blocks. $n_i$ is the number of COMs in the block with level $i$. Now the average correlation of the function will be just
\begin{align*}
\left<Q(L_B,t)\right> &=\frac{1}{N_B}  \sum _{i=1}^{N_B} Q(L_B,t).
\end{align*}
The four-point susceptibility for each block is :
\begin{align*}
 \chi_4^{\prime}(L_B,t)&= [\left<Q(L_B,t)^2\right>-\left<Q(L_B,t)\right>^2].
\end{align*}
So the averaged susceptibility will be 
\begin{align*}
\chi_4(L_B,t) &=\frac{N_R}{N_B} \sum _{i=1}^{N_B} \chi_4^{\prime}(L_B,t).
\end{align*}
where $\left< \cdots \right>$ denotes averages over different grand canonical ensembles of size $L_B$.

\subsection{Spatial correlation of mobility :}
The four-point structure factor can be written as :
\begin{equation*}
    S_4(\vec{q},t)=\frac{1}{N}(\langle\tilde{Q}(\vec{q},t)\tilde{Q}(-\vec{q},t)\rangle-\langle\tilde{Q}(\vec{q},t)\rangle\langle\tilde{Q}(-\vec{q},t)\rangle)
\end{equation*}
where, 
\begin{equation*}
\tilde{Q}(\vec{q},t)=\sum_{i=1}^{N_R}e^{i\vec{q}.\vec{r}_{i,COM}(0)}w(|\vec{r}_{i,COM}(t)-\vec{r}_{i,COM}(0)|)
\end{equation*}
We can fit $S_4(\vec{q},\tau_{\alpha})$ to (for small values of q) from the Ornstein-Zernike form:
\begin{equation*}
    S_4(q,\tau_\alpha)=\frac{S_4(0,\tau_\alpha)}{(1+q^2\zeta^2)}
\end{equation*}
where $\zeta$ is the dynamic length scale.

\section{The effect of changing number of monomers in each ring on the relaxation time and dynamic susceptibility :}
\begin{figure}[!htb]
  \raggedright
  \includegraphics[width=1.00\columnwidth]{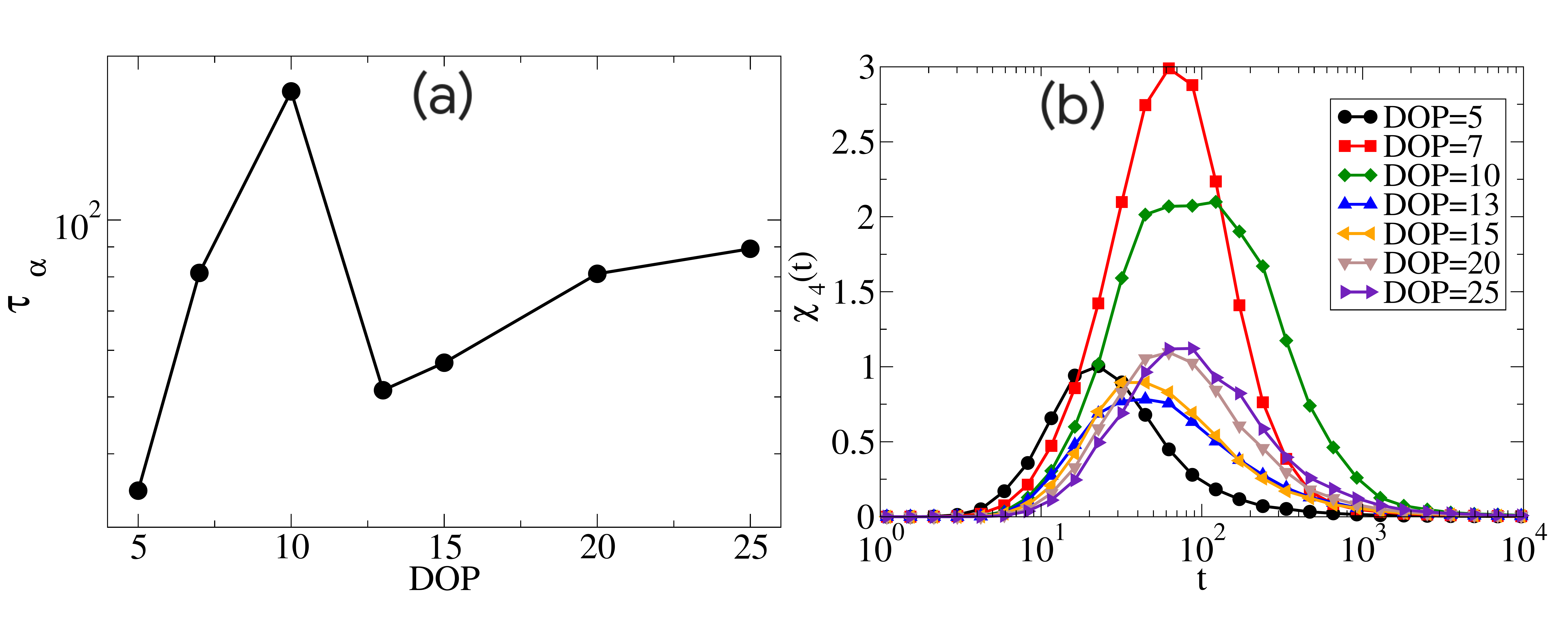}
  \caption{(a)The re-entrant relaxation times with changing the number of monomers in each ring. (monomer density and the number of rings are fixed( $\rho=0.23$, Number of rings=289))(b) The four-point-suseptibility also shows the same trend of re-entrance with number of monomers.}
  \label{DOP}
\end{figure}

To study the effect of changing the number of monomers, we kept the equilibrium angle at $180^\circ$. So, it takes more force to deform the rings if the number of monomers are less. To keep the average environment fixed, we kept the number density and the number of rings fixed ( so the number of monomers has changed). We see a re-entrant behaviour in the relaxation times across changing the number of monomers in each ring. As the distance between the monomers are fixed, we expect if the number of monomers in each ring increases it would take a longer time fo the ring to diffuse. Also due to the deformability in the rings at a given density, there will be a sweet spot of number of monomers in each ring where the rings will be arranged in a more hexatic phase order than others. These two competing effects might give rise to a re-entrant like behaviour in the relaxation-time profile versus the degree of polymerization.( see \ref{DOP})

\section{Pronounced re-entrant behaviour at lower temperature :}
\begin{figure}[!htb]
  \raggedright
  \includegraphics[width=0.85\columnwidth]{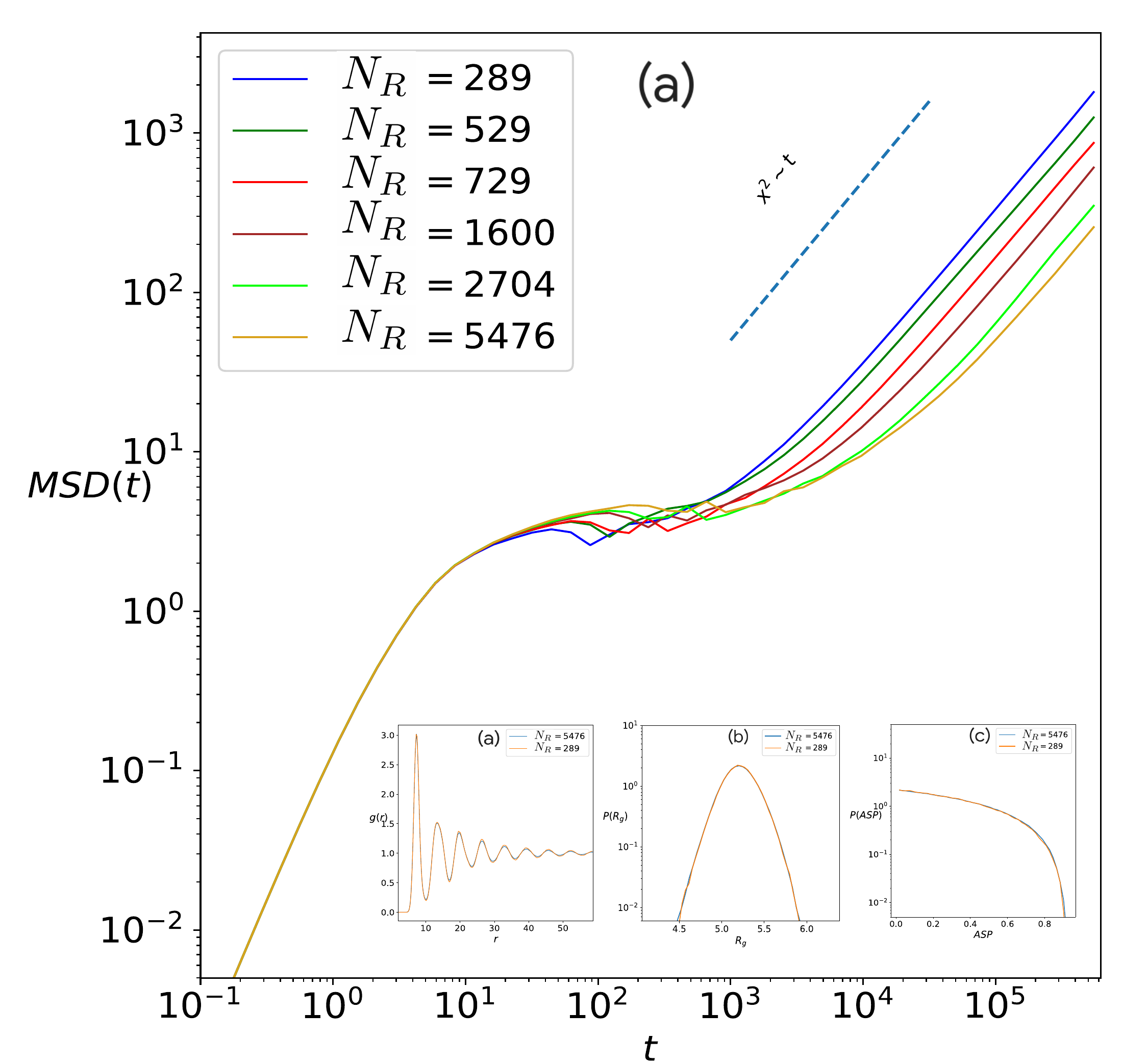}
  \caption{(a)The MSD plots are for $\rho=0.21$, $k_{\theta}=10.0$ and T=0.80, for the same system parameters (Inset shows the comparison of the static properties which makes the origin of the re-entrant behaviour entirely dynamic.)}
  \label{lowmsd}
\end{figure}

The re-entrant behaviour with system size increased with lowering temperature and density, and the mean-squared displacement showed the re-entrant behaviour till very large system sizes. Note that we didn't see any difference in the static factors as shown in the inset. We also checked the ASP and RG for a tagged ring and nothing significant was found.( see FIG. \ref{lowmsd}).

\begin{figure}[!htb]
  \raggedright
  \includegraphics[width=0.85\columnwidth]{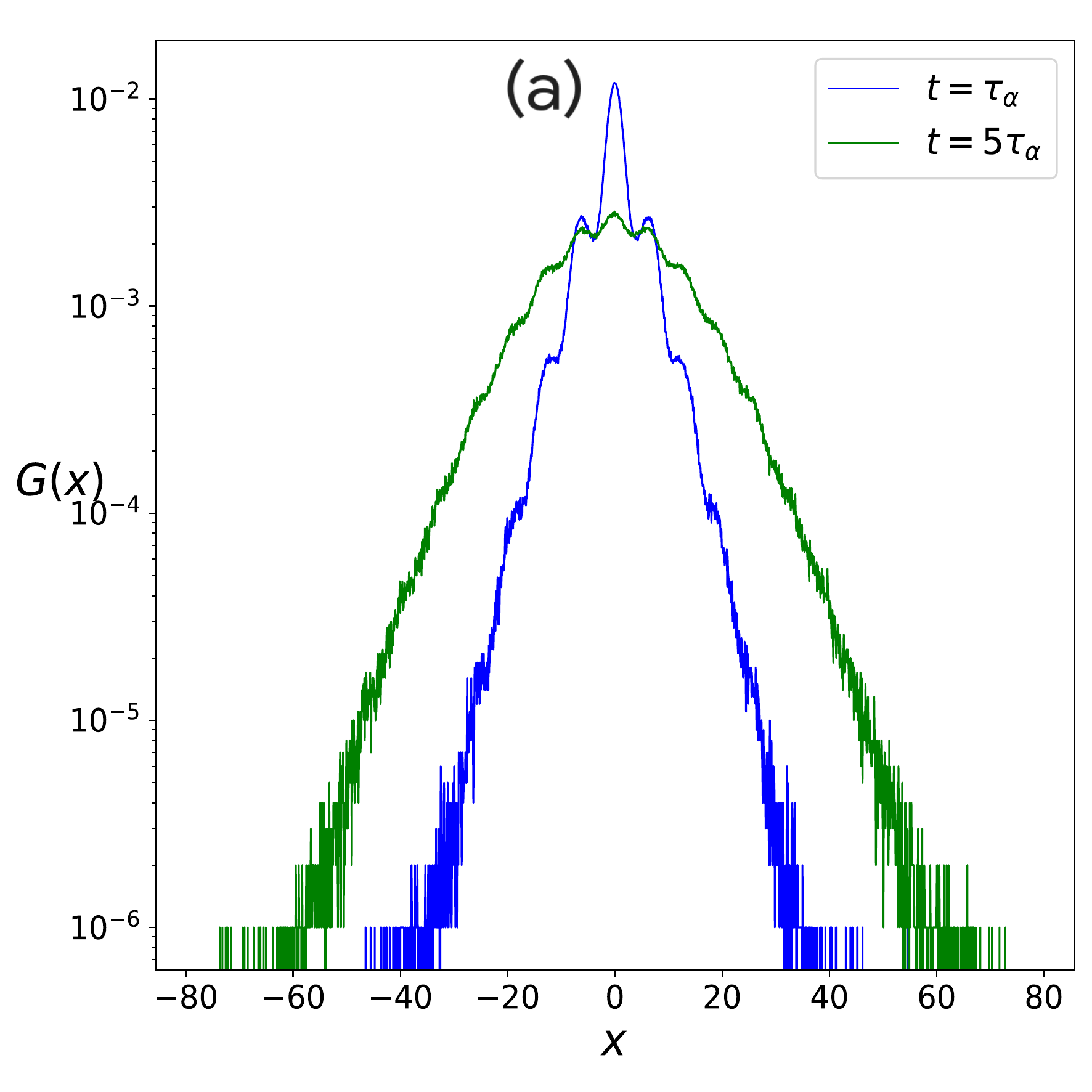}
  \caption{(a)The self part of the Van-Hove function at T=0.80 ($N_R$=289)}
  \label{VHLT}
\end{figure}

At this temperature and for this density hopping mechanism of movement of the rings between cages are prominent( see FIG. \ref{VHLT}), which are seen from shoulders arising in van-hove function, separated by the $x\approx6.5$. This basically means the hops are mostly of the same distance at this temperature. The time-evolved van-hove function has lower peaks which justifies the diffusion at longer times. We also checked the distance-overlap function( $d_\tau(t)=0$ if $|\vec{r_{i,COM}(t+\tau)}-\vec{r_{i,COM}(t)}|<a'$ and $d_\tau(t)=1$ if $|\vec{r_{i,COM}(t+\tau)}-\vec{r_{i,COM}(t)}|>a'$)( we chose $a'\approx R_g$) averaged over all the rings at these temperatures for the soft particles to check for any intermittency from a jammed to fluid behaviour in these systems and the overlap-function turned out to be fairly fluid-like. 

\section{The variation of the VFT fits across the re-entrant system sizes :}

\begin{figure}[!htb]
  \raggedright
  \includegraphics[width=0.85\columnwidth]{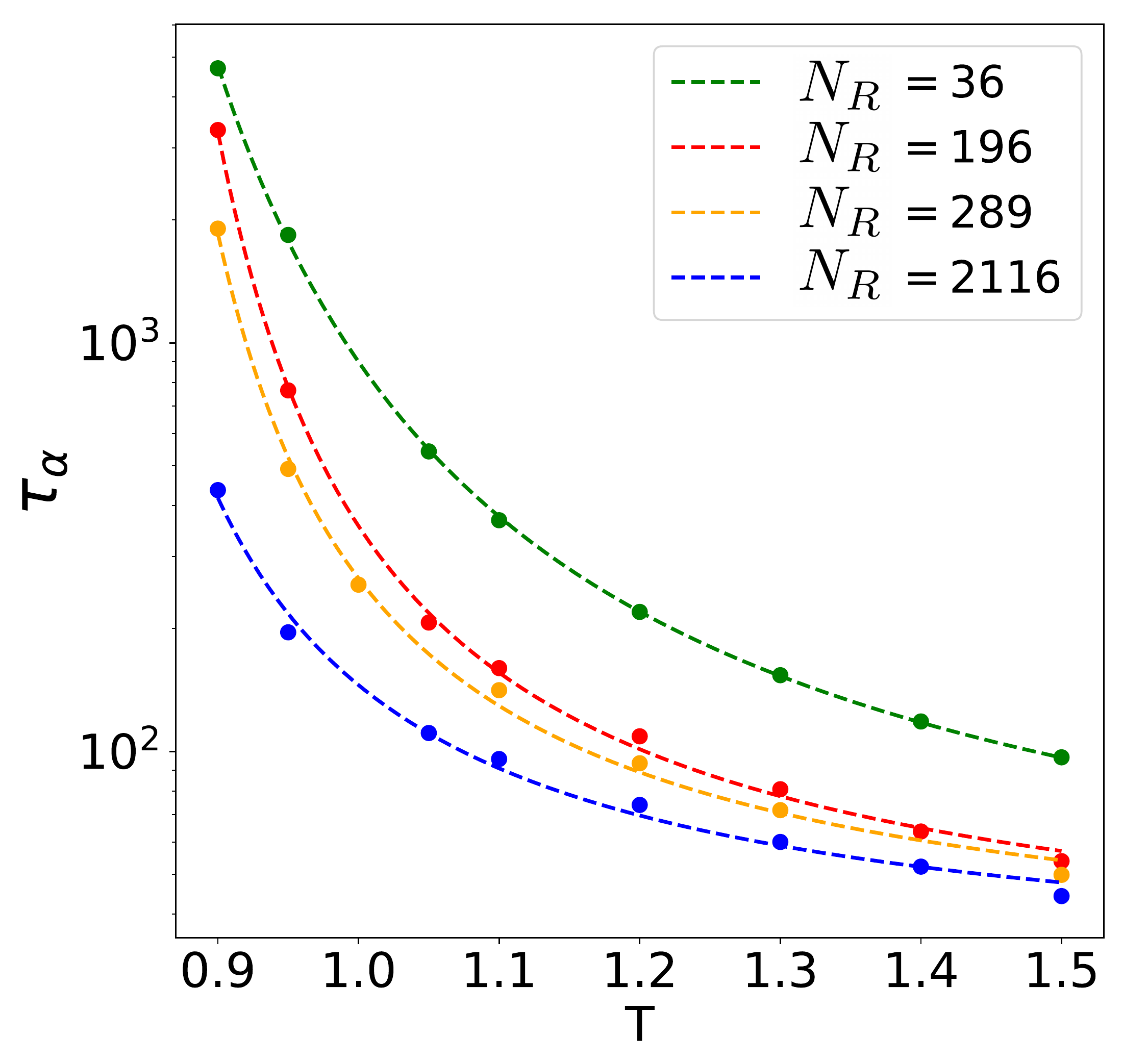}
  \caption{The VFT fits for different number of rings for $k_{\theta}=10.0$.}
  \label{VFT}
\end{figure}

We fitted the data of the relaxation times for different system sizes for $k_{\theta}=10.0$ for different temperatures across the re-entrant system sizes. The variation of the VFT-parameters were $K_{VFT}=0.584-1.679$ ( where the smallest value is for the smallest system size $N_R$=36), $\tau_0=23.27-27.15$, and $T_{VFT}=0.681-0.783$. (see \ref{VFT})

\section{Suppression of Mermim-Wagner like fluctuations via cage-relative MSD for higher temperatures :}

\begin{figure}[!htb]
  \raggedright
  \includegraphics[width=1.00\columnwidth]{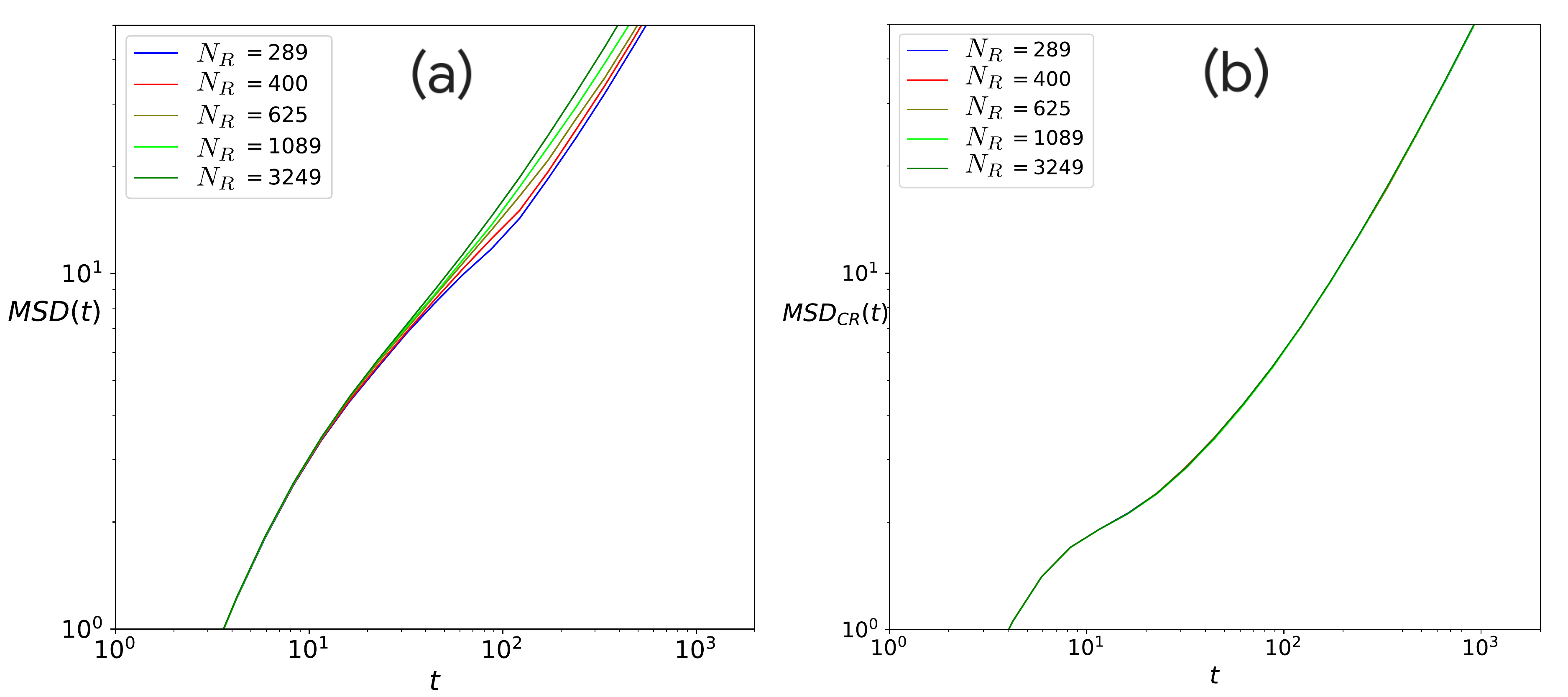}
  \caption{(a)The actual MSD plots with system sizes, (b)The cage-relative MSD plots with system sizes which fall almost on top of each other due to suppression of the Mermim-Wagner effects.}
  \label{CR}
\end{figure}

Due to the nature of the cage-relative mean-squared displacement, where displacements are considered with respect to the particles' initial cages, as the sizes of the cages are the same we can find a suppression of the fluctuations in MSD due to Mermim-Wagner effect. This suppression is lost in the case of lower temperatures. ( see \ref{CR})

\section{Collapse of $\chi_{4P}$ with the dynamic length scale obtained :}

\begin{figure}[!htb]
  \raggedright
  \includegraphics[width=1.00\columnwidth]{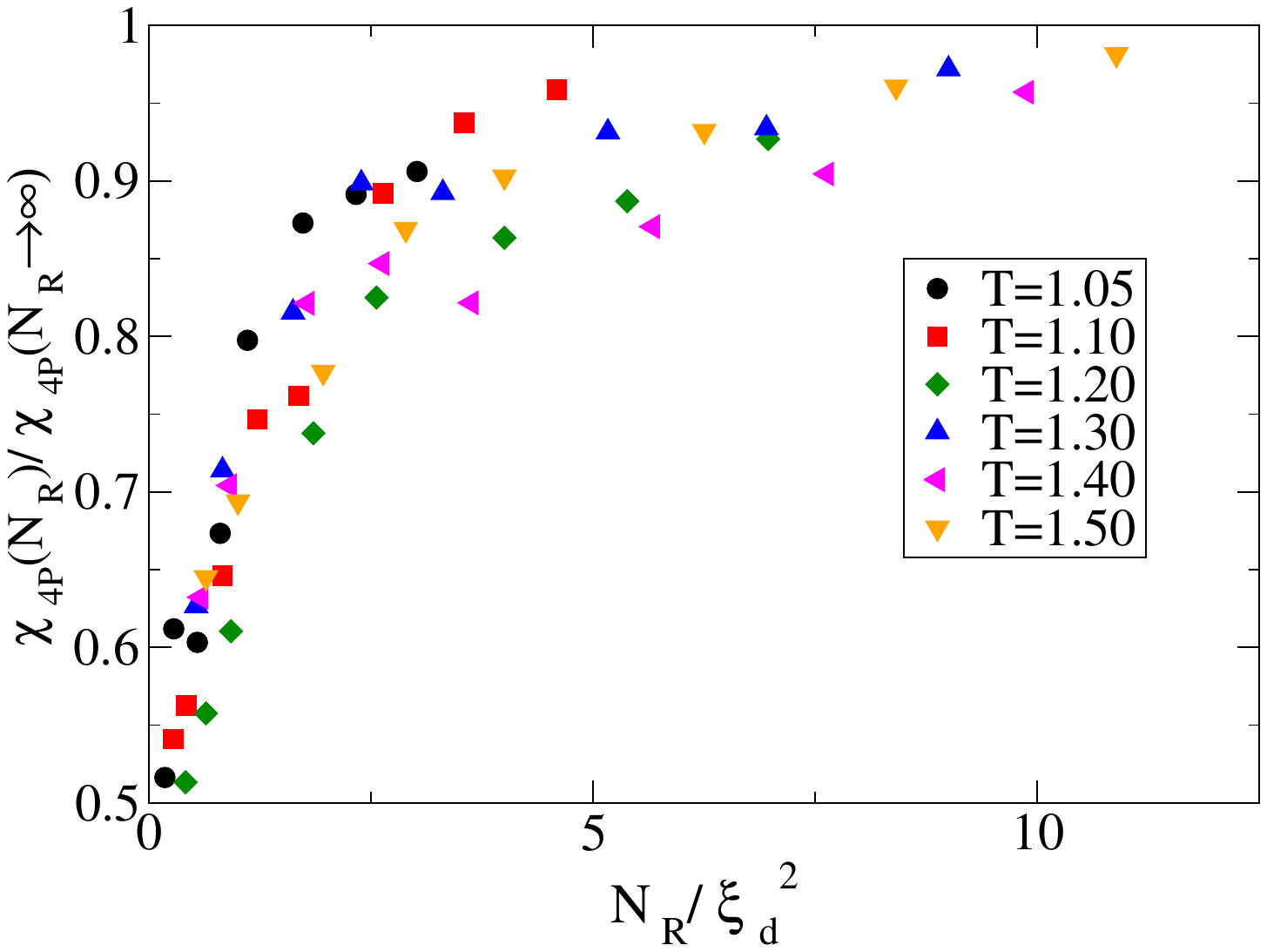}
  \caption{The collapse of scaled $\chi_{4P}$ vs the length scale-scaled number of rings}
  \label{chicoll}
\end{figure}

We expect the scaled-raw-$\chi_{4P}$ to collapse with the dynamic-length scales obtained in the main text. But due high amount of non-monotonicity of the $\chi_{4P}$ with $N_R$, which is systematic, a good collapse couldn't be found. The fits are not that good and it became worse when we tried to fit the scaled-$\chi_{4P}$s for lower temperatures, hence we give the collapse for the higher temperatures only.( see \ref{chicoll}) The fit form is similar to the block analysis i.e. $\chi_{4P} (N_R)/\chi_{4P} (N_R\to\infty)=f(N_R/\xi_d^2)$

\section{Lowering of Relaxation time( $\tau_{\alpha}$) at lower densities:}
\begin{figure}[!hptb]
  \includegraphics[width=0.90\columnwidth]{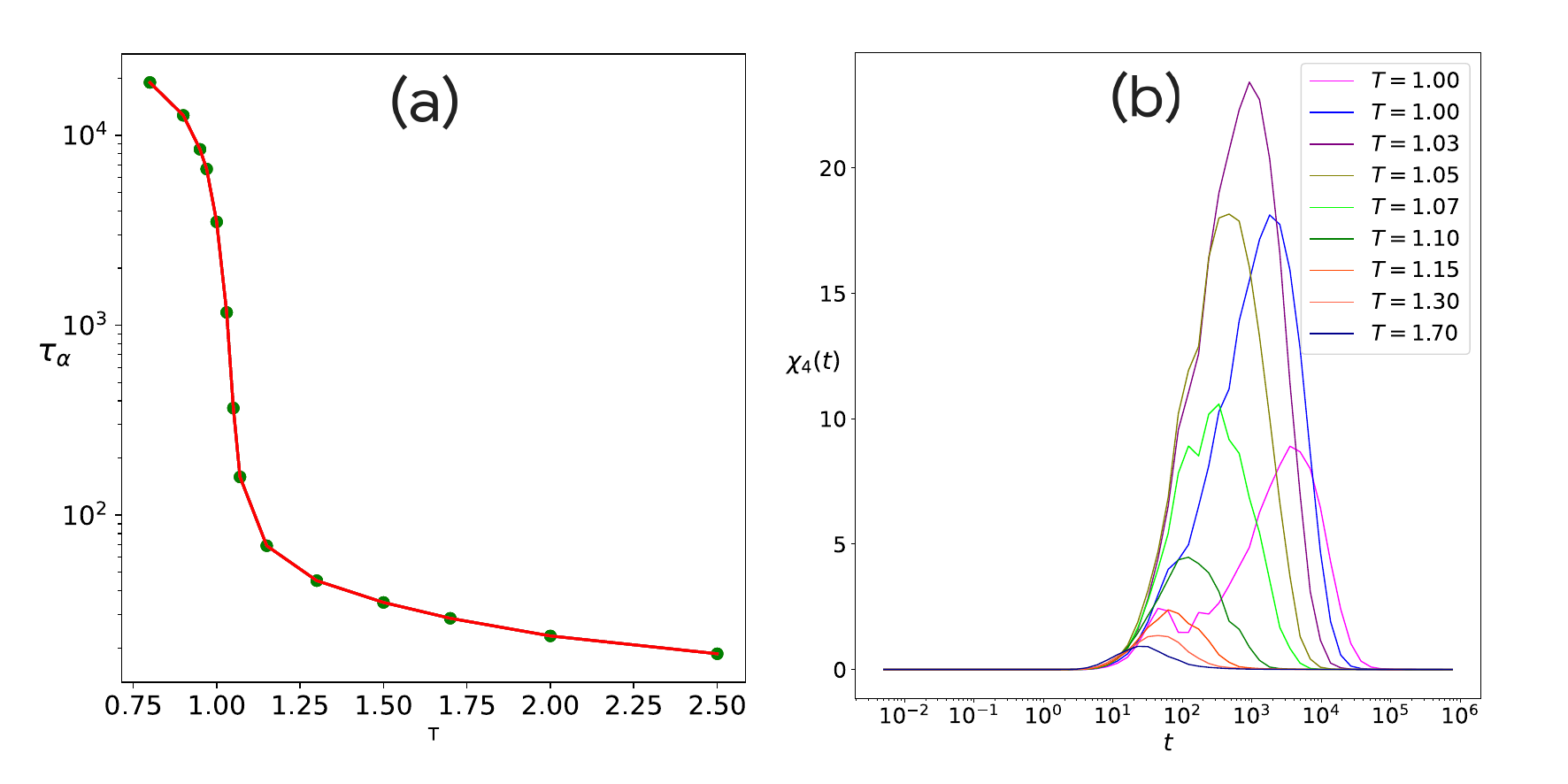}
  \caption{(a)The saturation of $\tau_{\alpha}s$ for smaller temperatures at lower densities. (b)The $\chi_4(t)$ plots vs time for different temperatures at same system parameters $\rho=0.21$ and $k_{\theta}=10.0$ ($N_R$=289)}
  \label{tausat}
\end{figure}

At lower density( $\rho=0.21$) we get a saturation of $\tau_{\alpha}$ for lower temperatures( see FIG. \ref{tausat}). At lower densities the system behaves in a more fluid-like way, which makes the system relax quickly as expected. A saturation with temperature behaviour is also seen.

\section{Van Hove function and its time evolution:}
\begin{figure}[!hptb]
  \includegraphics[width=0.90\columnwidth]{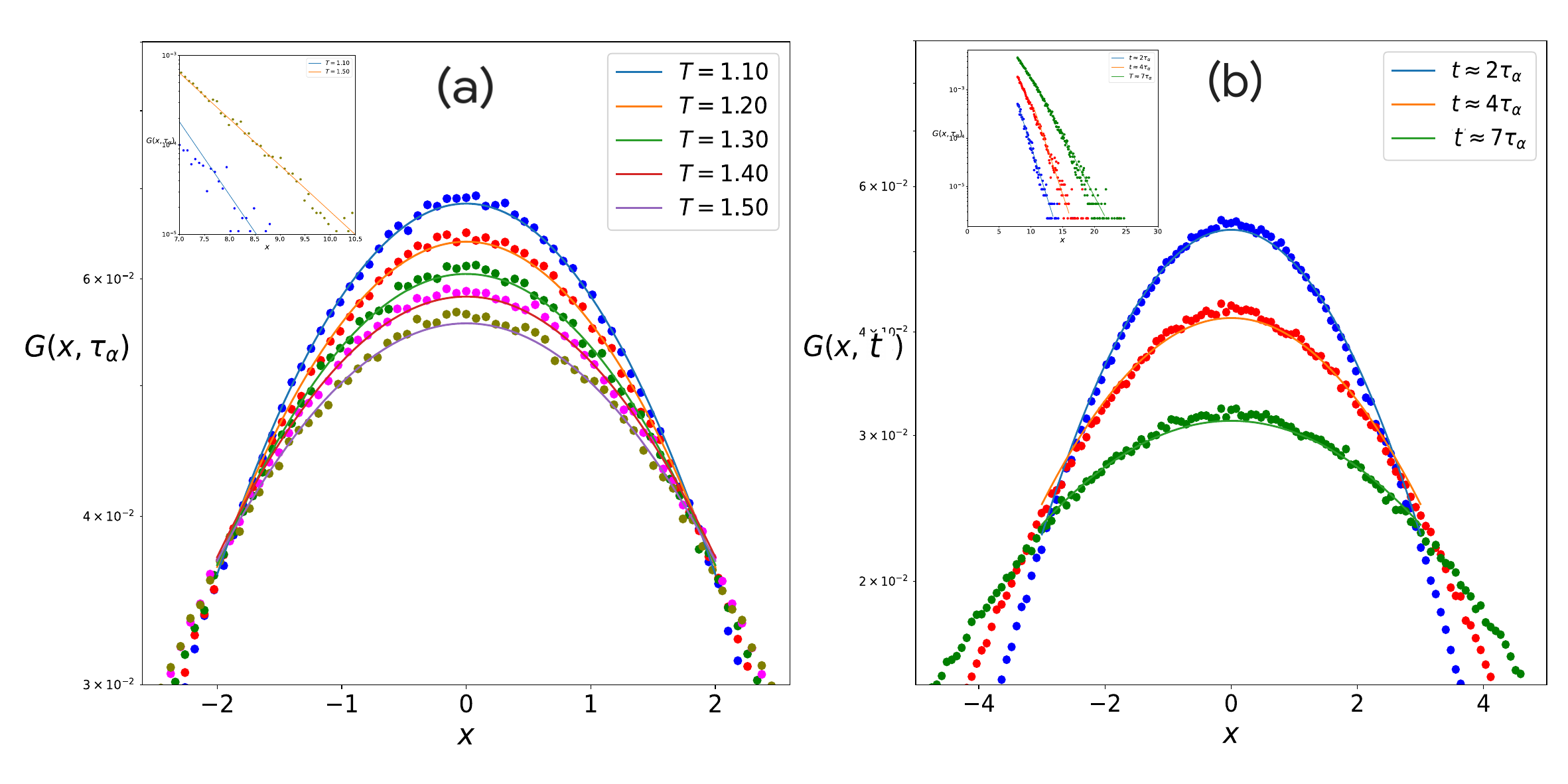}
  \caption{\small (a)Gaussian fits at different temperatures to the Van hove functions, inset shows the exponential decay at large r (b) Gaussian fits at different times for T=1.05 to the Van hove functions, inset shows the exponential fits ($N_R$=289)}
  \label{Vanhove}
\end{figure}

The Van hove functions is known to have a gaussian centre and an exponential tail ar large r, which happens due to dynamic heterogeneity in a system\cite{PhysRevLett.99.060604}. The Van hove function when plotted at the $t\approx\tau_{\alpha}$, shows peak heights that are higher for lower temperature as the particles have not moved that far( see FIG. \ref{Vanhove}). As, time evolves the peak height decreases as more and more particles move out of their cage.

The decrease in peak height with temperature is again a sign of decrease in the dynamical heterogeneity length scale in the system.

%\newpage

\section{Langevin dynamics simulations :}
\begin{figure}[!hptb]
  \includegraphics[width=0.9\columnwidth]{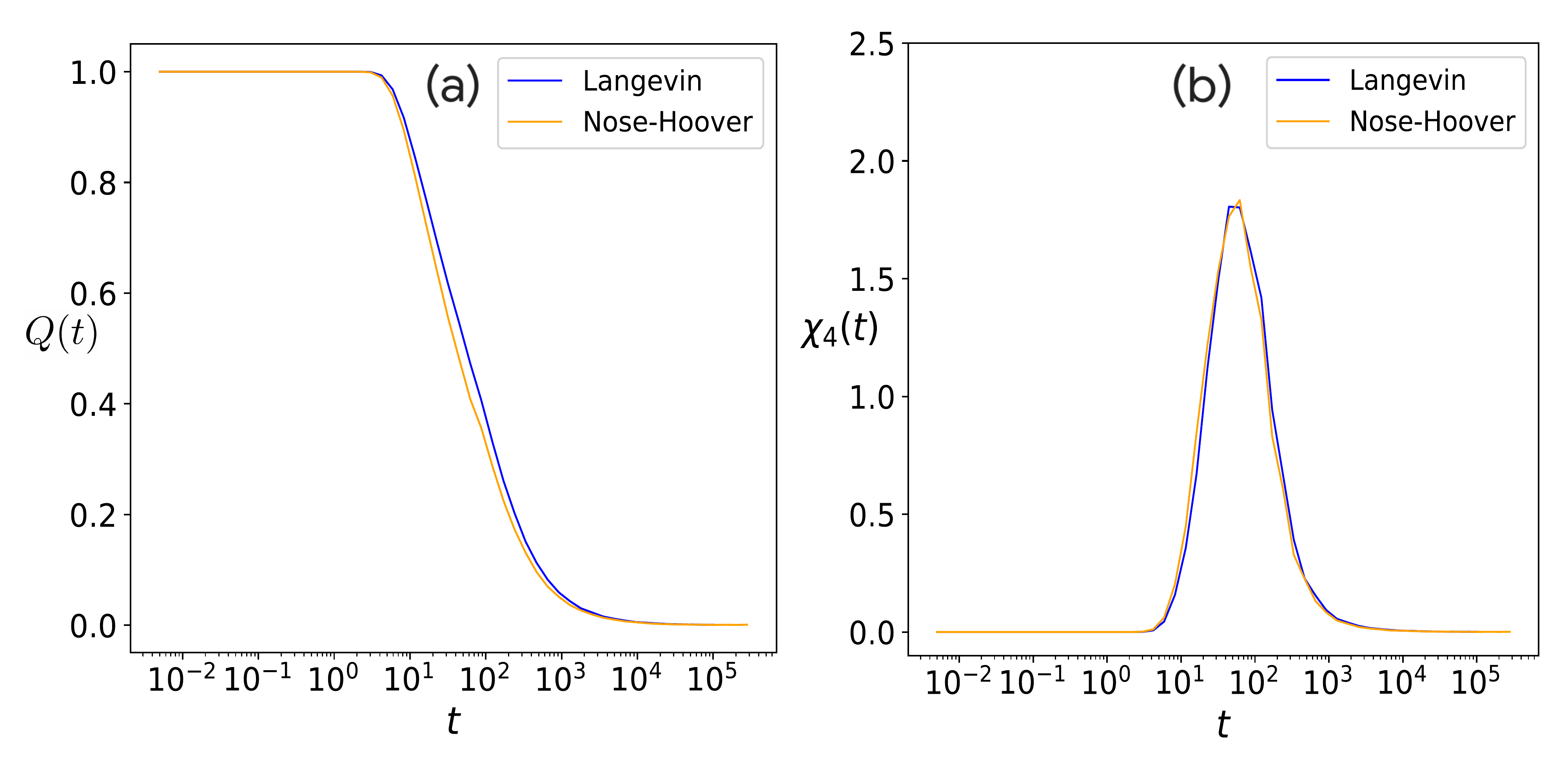}
  \caption{\small(a)The overlap function (Q(t)) as a function of time  (b)The dynamic susceptibility ($\chi_4(t)$) as a function of time}
  \label{lannose}
\end{figure}

We also tried langevin dynamics simulations for checking the robustness of our results and the alpha-relaxation time( $\tau_\alpha$) was almost the same, and also the dynamic susceptibility( $\chi_4(t)$) was the same with the damping parameter(=50.0). ( see FIG. \ref{lannose})

\section{Suppression of dynamic heterogeneity across density :}
\begin{figure}[!htb]
  \includegraphics[width=0.90\columnwidth]{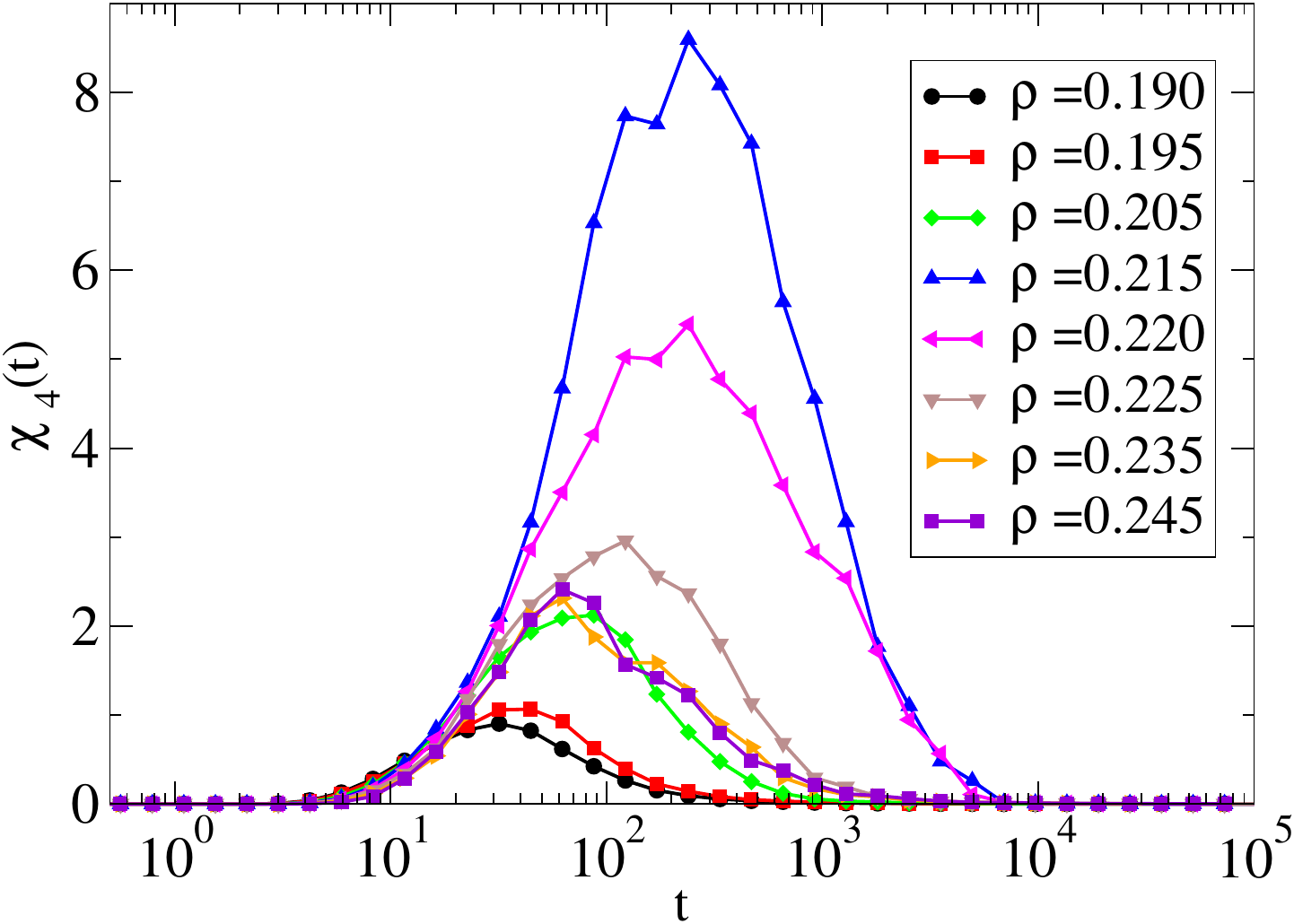}
  \caption{\small The suppression of peak values of the dynamic susceptibility across densities.( for $k_\theta=15.0$)}
  \label{SupofDHacrossdensity}
\end{figure}

The suppression of dynamic heterogeneity across densities. Close to the jamming transition the dynamic heterogeneities gets suppressed in colloids, as has been reported experimentally in \cite{Ballesta_2008}. The non-gaussianity parameter( $\alpha_2(t)$) also shows a similar trend as the four-point susceptibility.  We see a similar trend here together with the re-entrant behaviour in the relaxation times.( see Fig. \ref{SupofDHacrossdensity})

\section{Compensatory effects of temperature and stiffness in the dynamics :}
\begin{figure}[!htb]
  \raggedright
  \includegraphics[width=1.00\columnwidth]{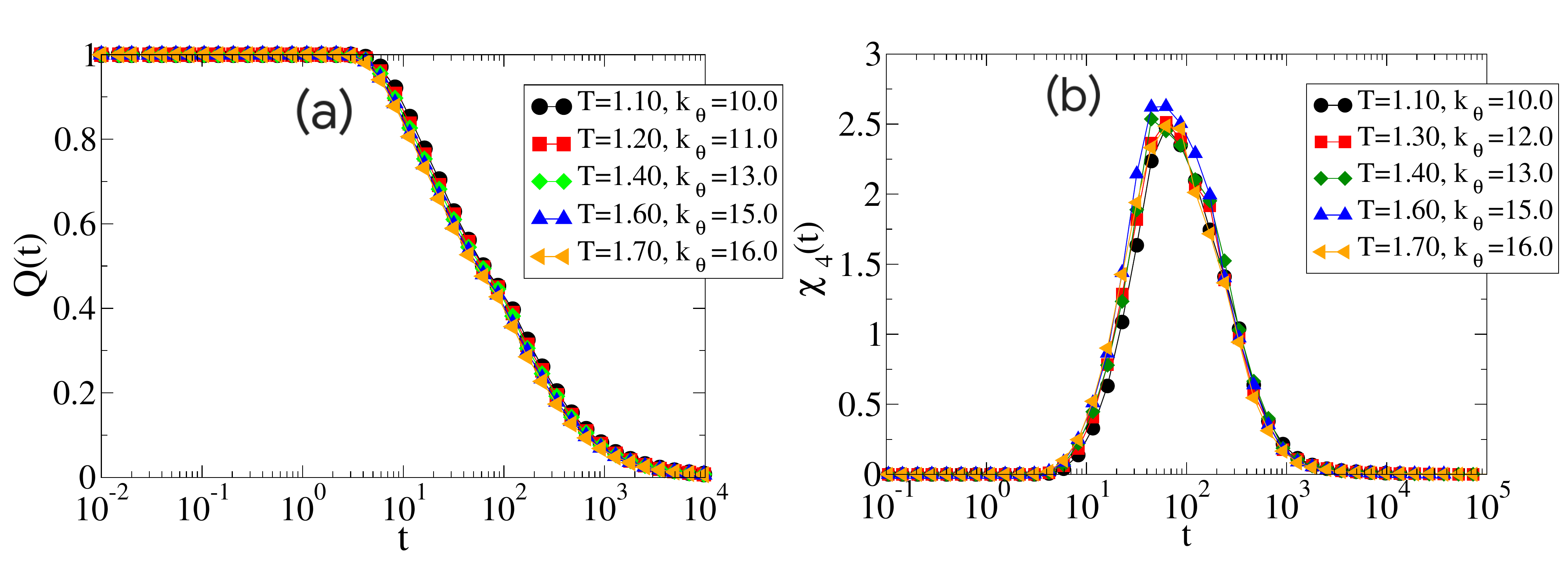}
  \caption{\small (a) Overlap functions for various values of stiffness( $k_{\theta}$) and temperature, which fall on top each other. ( Note the small but systematic deviation of the ovelap function for higher temperatures.)(b) The four-point susceptibility for these combinations also fall on top of each other and there is no enhancement or suppression of dynamic-heterogeneity due to softness.}
  \label{compen}
\end{figure}

We expect the effect of increasing the stiffness( increasing $k_{\theta}$) and decreasing the temperature to be the same in the dynamics of the rings. So, we expect some combinations of $k_{\theta}s$ and temperatures to give similar dynamics. One such combination is shown in the plots below. We see that both the overlap function and the four-point susceptibility function fall almost on top of each other. The compariosn is for $N_R=289$. ( see Fig. \ref{compen})

\section{Static factors :}

\begin{figure}[!hptb]
  \raggedright
  \includegraphics[width=1.00\columnwidth]{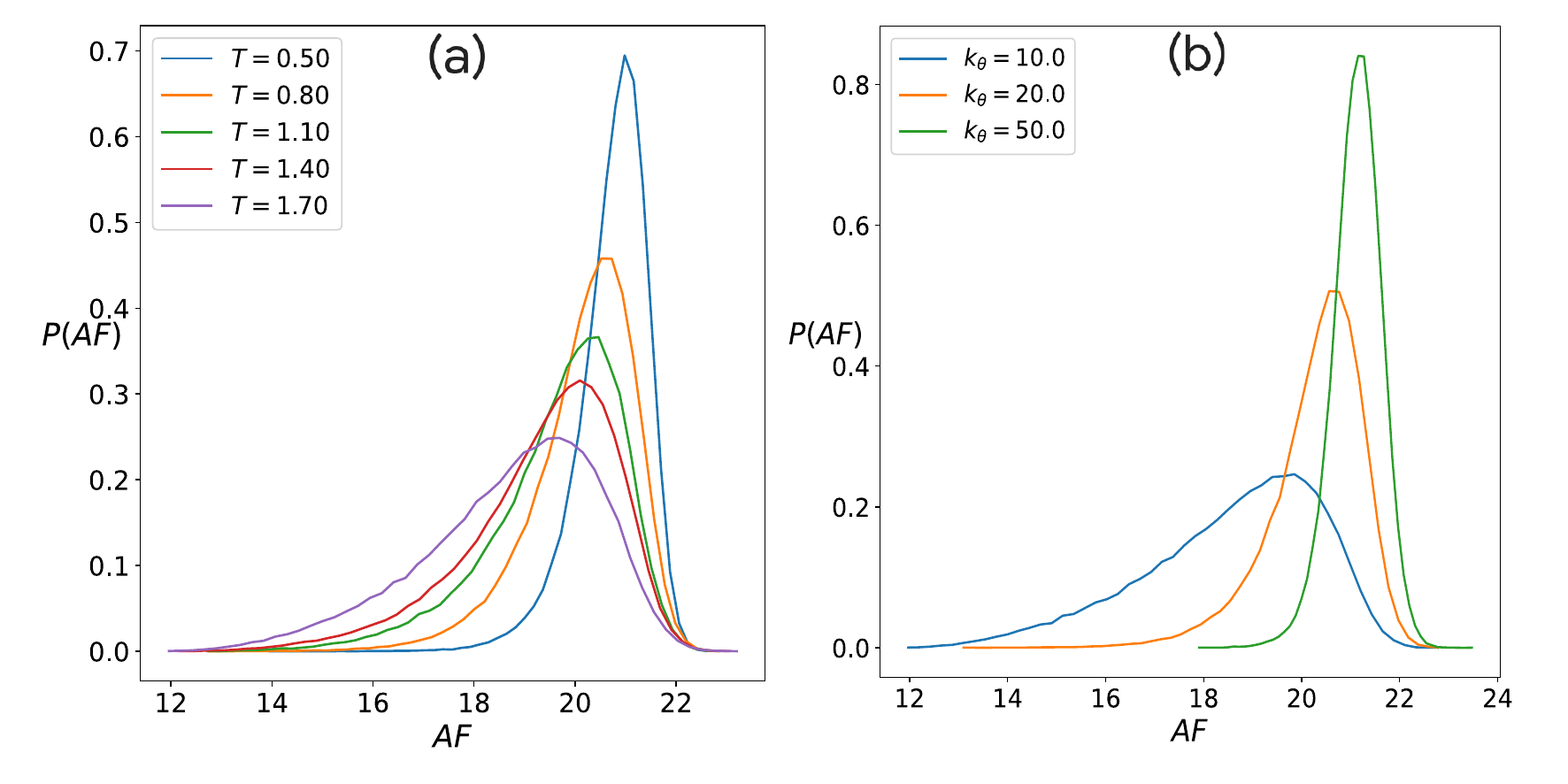}
  \caption{\small (a)The plots of P(AF) vs AF for different temperature at $k_{\theta}=10.0$ (b)The plots of P(AF) vs AF for different $k_{\theta}$ at $T=1.0$ ($\rho=0.23$) ($N_R=289$)}
  \label{AF}
\end{figure}

We plot the distribution of the area fraction (AF) for different stiffnesses and temperatures (see Fig. \ref{AF}). Again there is a similarity between increasing temperature and decreasing stiffness constant. 

\newpage

\bibliographystyle{plain}  % You can use other styles like alpha, abbrv, ieeetr, etc.
\bibliography{mainsup}